\documentclass[a4paper,11pt]{article}
\pdfoutput=1 

\usepackage{jcappub} 
\usepackage[T1]{fontenc} 
\usepackage[utf8]{inputenc}
\usepackage{graphicx}
\usepackage{epsfig}
\usepackage{rotate}
\usepackage{amsmath}
\usepackage{amssymb}
\usepackage{amsfonts}
\usepackage{bm}
\usepackage{enumerate}
\usepackage{enumitem}
\usepackage{afterpage}
\usepackage{xcolor}
\usepackage{multirow}
\usepackage{array}
\usepackage{natbib, hyperref}
\usepackage[normalem]{ulem}

\newcommand{\ud}{\mathrm{d}}

\newcommand{\De}{\Delta}
\newcommand{\cH}{\mathcal{H}}

\newcommand{\OO}{\mathcal{O}}
\newcommand{\bn}{\bm{n}}
\newcommand{\h}{{\rm HI}}
\newcommand{\np} {\bm{\nabla}_\perp}
\def\be{\begin{equation}}
\def\ee{\end{equation}}
\def\bea{\begin{eqnarray}}
\def\eea{\end{eqnarray}}

\definecolor{dgreen}{rgb}{0,0.7,0.0}

\title{Full-sky bispectrum in redshift space for 21cm intensity maps}

\author{{Ruth} Durrer$^1$, {Mona} Jalilvand$^1$,  {Rahul} Kothari$^2$, \\ {Roy} Maartens$^{2,3}$, {Francesco} Montanari$^4$}
\affiliation{$^{1}$Departement de Physique Th\'eorique, Universit\'e de Gen\`eve, 1211 Gen\`eve 4, Switzerland \\
$^2$Department of Physics \& Astronomy, University of the Western Cape, Cape Town 7535, South Africa\\
$^3$Institute of Cosmology \& Gravitation, University of Portsmouth, Portsmouth PO1 3FX, UK \\
$^4$ Instituto de F\'isica Te\'orica IFT-UAM/CSIC, Universidad
  Aut\'onoma de Madrid, Cantoblanco 28049 Madrid, Spain
}

\emailAdd{ruth.durrer@unige.ch}
\emailAdd{mona.jalilvand@unige.ch}
\emailAdd{quantummechanicskothari@gmail.com}
\emailAdd{roy.maartens@gmail.com}
\emailAdd{francesco.montanari@uam.es}

\abstract{
{We compute the tree-level bispectrum of 21cm intensity mapping after reionisation. We work in directly observable angular and redshift space, focusing on equal-redshift correlations and thin redshift bins, for which the lensing contribution is negligible. We demonstrate the importance of the contributions from redshift-space distortions which typically dominate the result. 
{Taking into account the effects of telescope beams and foreground cleaning,}
we estimate the signal to noise,  and show that the bispectrum is detectable by both SKA in single-dish mode and HIRAX in interferometer mode, especially at the lower redshifts in their respective ranges.} 
}

\begin{document}
\maketitle
\flushbottom

\section{Introduction}
\label{sec:intro}

In the  era of precision cosmology, cosmological models can be robustly tested against the data, from cosmic  microwave background (CMB) surveys and surveys of the large-scale structure. An advantage of large-scale structure surveys is that the data is 3D and therefore potentially {contains} a lot more information. In addition to  surveys using galaxy number counts, there are surveys that measure the integrated spectral line emission from galaxies \cite{Fonseca:2016qqw,Kovetz:2017agg}. Since they do not attempt to resolve individual galaxies, such intensity mapping surveys can cover large volumes rapidly -- although they face the problem of foreground removal. The 21cm line of neutral hydrogen (HI)  is particularly important, since {hydrogen} is the most abundant element in the Universe. After reionisation, 21cm maps provide a {biased} tracer of the matter distribution and are a potentially powerful probe of cosmological models \cite{Hall:2012wd,Bull:2014rha,Bacon:2018dui,Fonseca:2019qek}.

HI intensity mapping has poor angular resolution but exquisite redshift accuracy. At each redshift selected within the telescope band, the map of brightness temperature $T_\h(z,\bn)$, where $\bn$ is {a unit direction from source to observer}, is akin to the CMB temperature map. Like the CMB map, the observed HI temperature contrast, $\Delta_\h=\Delta T_\h/\langle T_\h\rangle\equiv \Delta$, is not affected by lensing at first order \cite{Hall:2012wd,Alonso:2015uua,Fonseca:2015laa} so that
$\Delta^{L(1)}=  \Delta^{(1)}$, where 
\bea \label{d1}
\Delta^{(1)}(z,\bm n)=\delta_{\rm HI}^{(1)}(z,\bm n){+}\frac{1}{\cal H} \partial_r^2\,V^{(1)}(z,\bm n) \, .
\eea
Here $r$ is the comoving line-of-sight distance and we neglect terms that are suppressed by ${\cal H}/k$ in Fourier space. The first-order velocity potential in the redshift-space distortion (RSD) term is defined by ${v}^{(1)}_i=\partial_i V^{(1)}$.

The lensed temperature contrast is related to the unlensed one by 
\be
\Delta^L(z,\bm n)= \Delta(z,\bm n+\bm{\nabla}_\perp\phi), 
\ee
where $\bm{\nabla}_\perp$ is  the gradient operator on the 2D screen space orthogonal to $\bm n$. The  lensing potential at first order is
\be \label{phi1}
\phi^{(1)}=-2\int_0^r \ud\tilde r\, \frac{(r-\tilde r)}{\tilde r r}\varphi^{(1)} \quad \mbox{where}\quad 2\varphi^{(1)}=\Phi^{(1)} + \Psi^{(1)}\,.
\ee
Here the metric potentials in Poisson gauge (neglecting vector and tensor modes) are given by: 
\be
ds^2=a^2\Big[ -\big(1+2\Psi\big)d\eta^2+ \big(1-2\Phi\big)d\bm{x}^2\Big].
\ee

At second order\footnote{We use the convention $X=X^{(1)}+X^{(2)}+\cdots$.}, the CMB temperature map is affected by lensing deflection -- and  likewise for HI brightness temperature,
 as shown by \cite{Umeh:2015gza,Jalivand:2018vfz}:
\bea
\Delta^L(z,\bm n) 
= \Delta^{(1)}(z,\bm n) + \Delta^{(2)}(z,\bm n)-\big\langle \Delta^{(2)}\big\rangle (z) +L^{(2)}(z,\bn)-\big\langle L^{(2)}\big\rangle (z)
\,,\label{dlen}
\label{eq:lens_contri}
\eea
where the lensing correction,
\be
L^{(2)}(z, \bn)=\nabla_\perp^a\phi(z,\bm n)\,\nabla_{\perp a}\Delta^{(1)}(z,\bm n),
\ee
is a coupling of the deflection angle $\bm{\nabla}_\perp\phi^{(1)}$ with the gradient of the observed temperature contrast, $\np\Delta^{(1)}$.
This leads to a lensing correction to the 1-loop HI power spectrum \cite{Umeh:2015gza,Jalivand:2018vfz}.

Since the tree-level bispectrum includes second-order perturbations, the HI bispectrum is affected by lensing. For  CMB, this is  not the case: the tree-level CMB bispectrum  has no lensing contribution in the case of Gaussian initial conditions. The reason is that there is effectively no correlation between the primary temperature fluctuations generated at $z\sim 1000$ and the lensing deflections induced by large-scale structure at $z\lesssim 10$. Since this correlation is not negligible for HI intensity, we expect the lensing contribution to the 21cm bispectrum to be nonzero at tree-level \cite{Jalivand:2018vfz}.
This was already shown in~\cite{DiDio:2015bua}. However, as was also shown there, at equal redshifts and for narrow redshift bins, the lensing terms are always several orders of magnitude smaller than the contributions from density and redshift space distortions. For this reason we neglect them in our numerical analysis where we concentrate on equal-redshift bins.
The lensed 3-point correlation function  is 
\begin{equation}
{B}^L(z_i,\bn_i)=\big\langle {\Delta}^L_1\,{\Delta}^L_2\,{\Delta}^L_3\big\rangle=\big\langle{\Delta}_1\,{\Delta}_2\,{\Delta}_3\big\rangle + \delta {B} \quad \mbox{where}~ \Delta_i = {\Delta}(z_i, \bm n_i)\,.
\end{equation}
At  tree-level, by \eqref{dlen} the lensing correction is 
\begin{equation}
\delta {B} = \big\langle{\Delta}^{(1)}_1 {\Delta}^{(1)}_2 \big[L^{(2)}_3-\big\langle L^{(2)}_3\big\rangle \big] \big\rangle + \mathrm{2\ perms}. \label{tree_level_lensing}
\end{equation}
By Wick's theorem, 
\begin{align}\label{tree_level_lensing_expansion} \big\langle \Delta_1^{(1)}\, \Delta_2^{(1)}\, L^{(2)}_3  \big\rangle &= \big\langle
\Delta_1^{(1)}\, \nabla_{\perp}^{a}\phi_{{3}}
\big\rangle \big\langle \Delta_2^{(1)}\,{\nabla_{\perp a}}\Delta_3^{(1)}  \big \rangle + 
\big\langle
\Delta_2^{(1)}\, \nabla_{\perp}^{a}\phi_{{3}}
\big\rangle \big\langle \Delta_1^{(1)}{\nabla_{\perp a}}\Delta_3^{(1)} \big\rangle  \nonumber\\
&~~~~ + \big\langle \Delta_1^{(1)} \Delta_2^{(1)}\big\rangle \big\langle \nabla_{\perp}^{a}\phi_{{3}}\,\nabla_{\perp a}\Delta_3^{(1)}  \big\rangle \, .
\end{align}
The first two terms in (\ref{tree_level_lensing_expansion}) {give non-vanishing contributions to the bispectrum (the vectors $\langle \Delta_1^{(1)}\,{\nabla_{\perp a}}\Delta_3^{(1)} \rangle$ and $\langle
\Delta_2^{(1)}\, \nabla_{\perp}^{a}\phi_{{3}}\rangle$ have directions defined {respectively} by the angle between $\bn_1$ and $\bn_3$ \& between $\bn_2$ and $\bn_3$),  while the third term cancels the second term of \eqref{tree_level_lensing}.}

Here our focus is on the tree-level bispectrum, with Gaussian primordial fluctuations and in equal redshift bins. For galaxy and 21cm surveys, the  angular bispectrum naturally includes both lensing effects and wide-angle correlations on the curved sky \cite{DiDio:2015bua,DiDio:2018unb} -- unlike the Fourier-space bispectrum \cite{Umeh:2016nuh}.  

Apart from RSD, the remaining `projection' effects from observing in redshift space are ultra-large scale relativistic effects, which arise from Doppler, Sachs-Wolfe, integrated SW and time-delay terms, and their cross-correlations with each other and the dominant density and RSD terms (at first order, see \cite{Yoo:2010ni,Challinor:2011bk,Bonvin:2011bg} and at second-order see \cite{Bertacca:2014dra,Yoo:2014sfa,DiDio:2014lka,Kehagias:2015tda,DiDio:2015bua,Umeh:2016nuh, Jolicoeur:2017nyt, Jolicoeur:2017eyi, Jolicoeur:2018blf,Clarkson:2018dwn,Maartens:2019yhx,deWeerd:2019cae}).  
These relativistic effects are all suppressed in Fourier space by factors $(\cH/k)^n$, where $n\geq {2}$ {in the power spectrum \cite{Yoo:2010ni,Challinor:2011bk,Bonvin:2011bg} and $n\geq 1$ in the bispectrum \cite{Umeh:2016nuh,Clarkson:2018dwn,Maartens:2019yhx}}, and we will neglect them.

The article is structured as follows. In Section \ref{sec:bispec_mul_space}  we derive the main bispectrum results, while in Section \ref{sec:SNR_Calcu} we present numerical calculations for the bispectrum and its signal to noise {ratio}, considering both single-dish and interferometer modes for future surveys with the SKA and HIRAX telescopes. {We conclude in Section~\ref{sec:conclusions}. 
We assume a fiducial flat $\Lambda$CDM cosmology, with $h=0.67, \Omega_{\rm b}=0.05, \Omega_{\rm cdm}=0.27, A_s=2.3 \times 10^{-9}, n_s=0.962, k_*=0.05/{\rm Mpc}$ for the Hubble constant,  baryon and cold dark matter density, amplitude, tilt and pivot scale of the primordial power spectrum, respectively.}
 
\section{HI angular bispectrum}
\label{sec:bispec_mul_space}

\begin{figure}[!ht]
\begin{centering}
\includegraphics[scale=0.45]{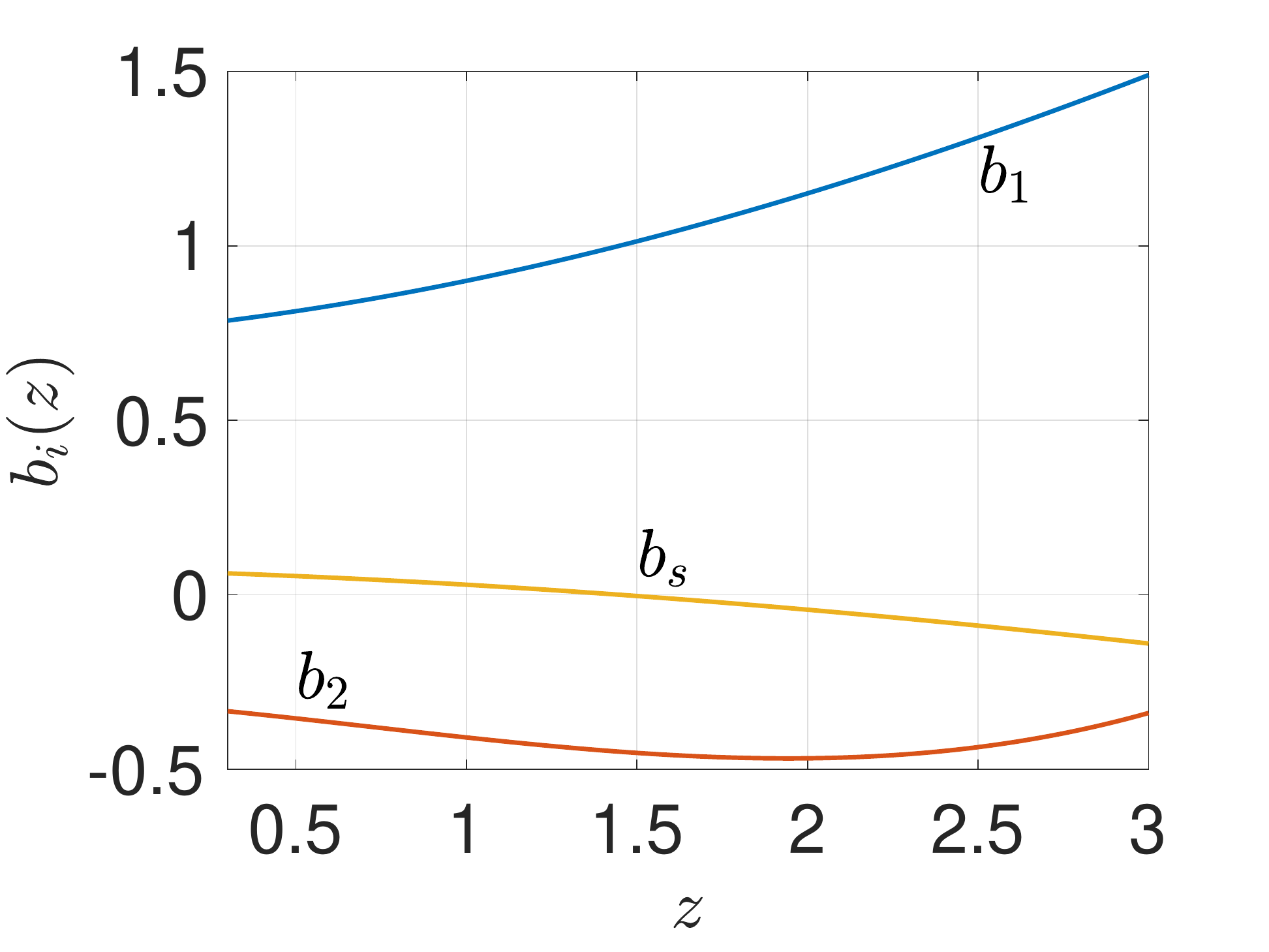}
\par\end{centering}
\caption{\label{fig:Bias_Coeff}HI bias coefficients \eqref{b1}--\eqref{bs}.}
\end{figure}

The (unlensed) HI temperature contrast is
\begin{equation}
{\Delta}(z,\bm{n})=\Delta^{(1)}(z,\bm{n})+{\Delta}^{(2)}(z,\bm{n})-\big\langle \Delta^{(2)}\big\rangle (z),
\end{equation}
where $\Delta^{(1)}$ is given by \eqref{d1}. Up to second order, using a standard bias model that includes tidal bias \cite{Desjacques:2016bnm}, and neglecting ultra-large scale relativistic effects, we have
\cite{DiDio:2018unb} {(see~\cite{Nielsen:2016ldx} for a simple, intuitive derivation of the second-order terms)}
\bea
\Delta^{(1)} &=&b_{1}\delta^{(1)}+\mathcal{H}^{-1}\partial_{r}^{2}V^{(1)}\\
\Delta^{(2)} & =&b_{1}\delta^{(2)}+\frac{1}{2}b_{2}\big[\delta^{(1)}\big]^{2}+b_{s}s^{2} +\mathcal{H}^{-1}\partial_{r}^{2}V^{(2)}
+ \mathcal{H}^{-2}\Big(\big[\partial_{r}^{2}V^{(1)}\big]^{2}+\partial_{r}V^{(1)}\partial_{r}^{3}V^{(1)}\Big),
\notag\\
&&{}+\mathcal{H}^{-1}\Big[\partial_{r}V^{(1)}\partial_{r} \delta^{(1)}+\partial_{r}^{2}V^{(1)} \delta^{(1)}\Big] .
\label{eq:delta_sec_ord}
\eea
Here $\delta$ is the  matter density contrast {in the comoving gauge, $V$ is the velocity perturbation in the Poisson gauge} and $s^2=s_{ij}s^{ij}$, where  the tidal field is
\begin{equation}
{s_{ij}=\Big(\partial_{i}\partial_{j}-\frac{1}{3}\delta_{ij}\nabla^2\Big)\nabla^{-2} \delta^{(1)}\,.}
\end{equation}
The bias coefficients can be modelled as \cite{Umeh:2015gza}
\bea
b_{1}(z) &=&~~0.754 +0.0877z +0.0607z^{2} -0.00274z^{3}\,,
\label{b1}\\
b_{2}(z)	 &=& -0.308-0.0724z -0.0534z^{2}+ 0.0247z^{3}\,, 
\label{b2}\\
b_{s}(z)	&=&-\frac{2}{7}\big[b_{1}(z)-1\big].
\label{bs}
\eea
{Note that \eqref{bs} is the simplest form of tidal bias, corresponding to zero tidal bias at the time of galaxy formation.}
Figure \ref{fig:Bias_Coeff} shows plots of these bias coefficients.

{Neglecting the lensing term which is subdominant at nearly equal redshifts,} the connected 3-point correlation function of the 21cm intensity map at tree level is\footnote{{Here and below, $\langle\Delta^{(2)}\rangle$ does not contribute to the connected part of the correlation function.}}
\begin{equation}
{B}(z_i,\bm{n}_i)=\big \langle \Delta^{(1)}_1\Delta^{(1)}_2{\Delta}_3^{(2)}\big\rangle + \mathrm{2 \ perms}.
\end{equation}
In angular harmonic space, 
\begin{equation}
{\Delta}(\bm{n},z)=\sum_{\ell m} {\Delta}_{\ell m}(z)Y_{\ell m}(\bm{n}), \label{eq:HI_Harm_Exp}
\end{equation}
so that
\begin{equation}
{B}(z_i,\bm{n}_i)=\sum_{\ell_i,m_i}B^{m_1m_2m_3}_{\ell_1\ell_2\ell_3}(z_1,z_2,z_3)Y_{\ell_1m_1}(\bm{n}_1)Y_{\ell_2m_2}(\bm{n}_2)Y_{\ell_3m_3}(\bm{n}_3),
\end{equation}
where 
\begin{equation}
B^{m_1m_2m_3}_{\ell_1\ell_2\ell_3}(z_1,z_2,z_3)={\big\langle {\Delta}^{(1)}_{\ell_1m_1}(z_1)\,{\Delta}^{(1)}_{\ell_2m_2}(z_2)\,{\Delta}^{(2)}_{\ell_3m_3}(z_3)\big\rangle + \mathrm{2 \ perms}},
\end{equation}
is the angular bispectrum.

On account of statistical isotropy,  ${B}(z_i,\bm{n}_i)$ can only depend on $\bm{n}_i\cdot\bm{n}_j$. This means that the $m_i$ dependence of the bispectrum takes the form
\begin{equation}
B^{m_1m_2m_3}_{\ell_1\ell_2\ell_3}(z_1,z_2,z_3)=\mathcal{G}^{m_1m_2m_3}_{\ell_1\ell_2\ell_3}\,b_{\ell_1\ell_2\ell_3}(z_1,z_2,z_3)\,
\end{equation}
where $\mathcal{G}^{m_1m_2m_3}_{\ell_1\ell_2\ell_3}$ is the Gaunt integral and $b_{\ell_1\ell_2\ell_3}$ is  the reduced angular bispectrum~\cite{Komatsu:2001rj}. The reduced bispectrum has contributions from the following six terms \cite{DiDio:2018unb}:
\begin{align}
b_{\ell_{1} \ell_{2} \ell_{3}}\left(z_{1}, z_{2}, z_{3}\right) &=b_{\ell_{1}\ell_{2}\ell_{3}}^{\delta^{(2)}}\left(z_{1}, z_{2}, z_{3}\right)+b_{\ell_{1} \ell_{2} \ell_{3}}^{v^{(2)^{\prime}}}\left(z_{1}, z_{2}, z_{3}\right)+b_{\ell_{2}^{\prime} \ell_{2} \ell_{3}}^{\delta v^{\prime}}\left(z_{1}, z_{2}, z_{3}\right) \nonumber\\ 
&+b_{\ell_{1} \ell_{2} \ell_{3}}^{v^{\prime 2}}\left(z_{1}, z_{2}, z_{3}\right)+b_{\ell_{1} \ell_{2} \ell_{3}}^{\delta^{\prime} v}\left(z_{1}, z_{2}, z_{3}\right)+b_{\ell_{1} \ell_{2} \ell_{3}}^{v^{\prime \prime}v}\left(z_{1}, z_{2}, z_{3}\right) \label{diff_bisp_cont}
\end{align}
The first term arises from the second order density contrast: {in Fourier space,} it contains monopole, dipole and quadrupole contributions (see~\cite{Bernardeau:2001qr,DiDio:2014lka} for details). The next term is the pure second order RSD contribution and the following four terms arise from the quadratic combinations of first-order  velocity and density perturbations appearing in \eqref{eq:delta_sec_ord}. 

\begin{figure}
\begin{centering}
\includegraphics[scale=0.45]{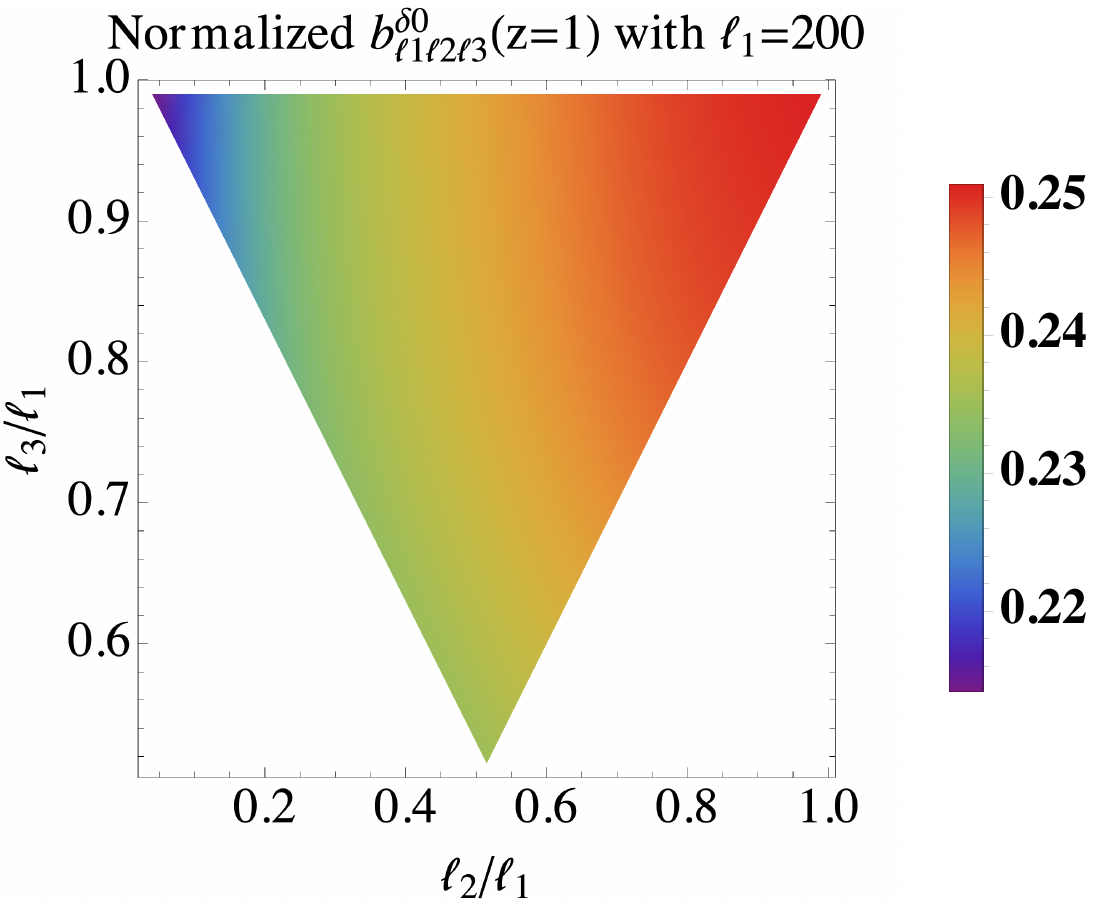}\includegraphics[scale=0.45]{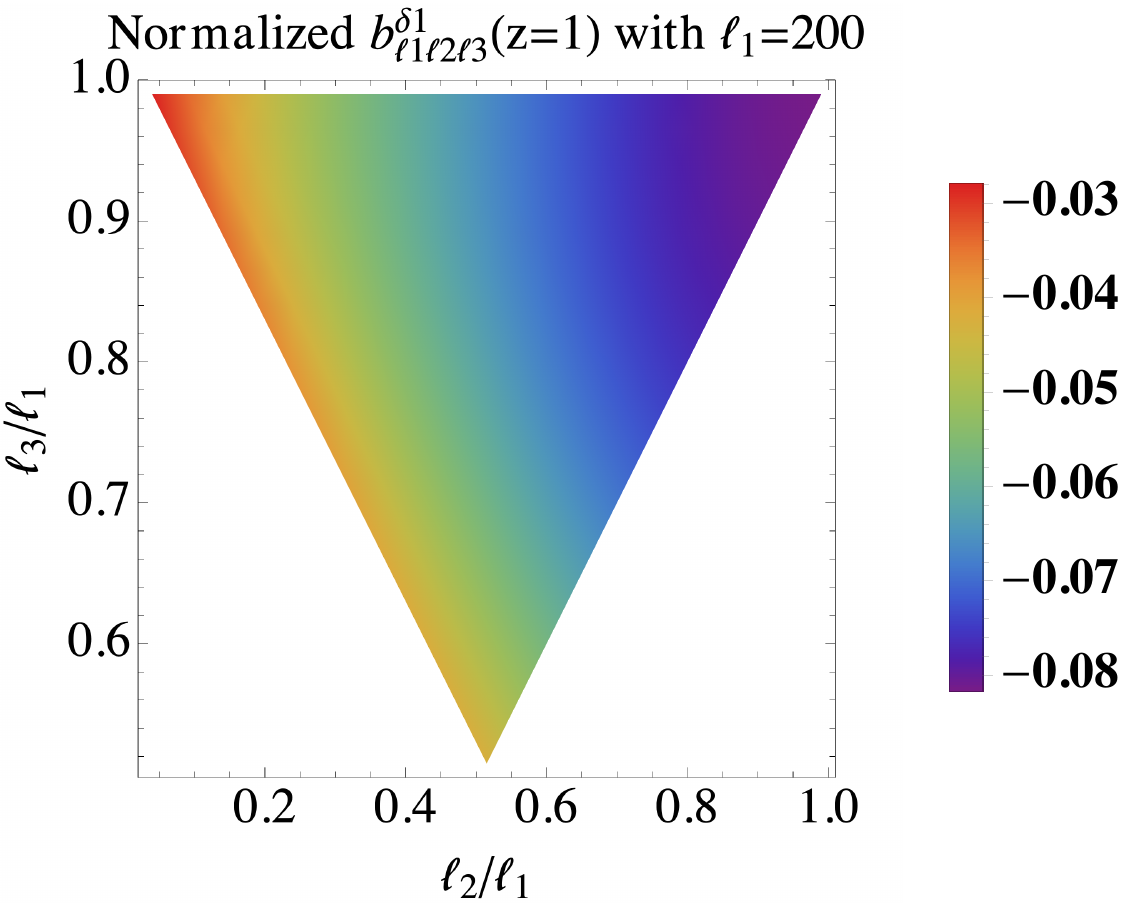}\includegraphics[scale=0.45]{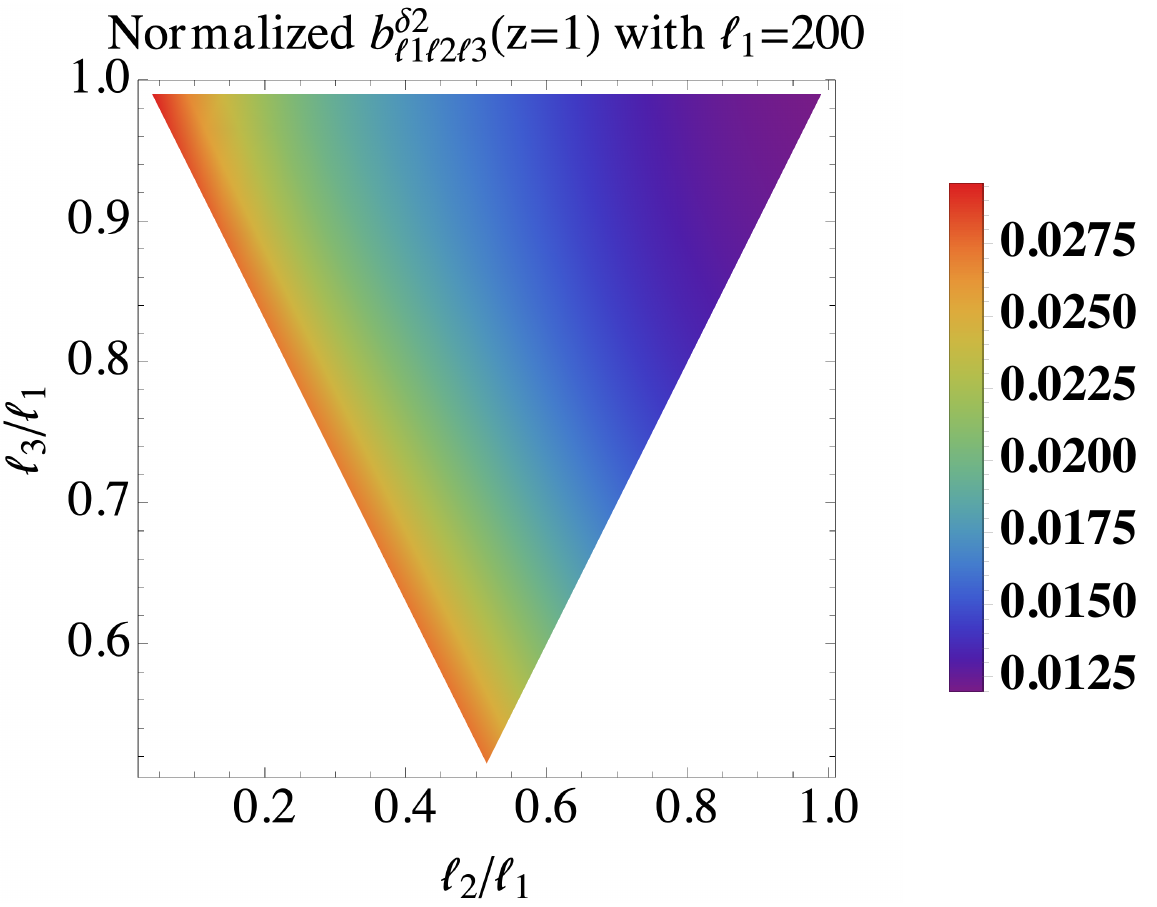}
\par\end{centering}
\caption{\label{fig:Den_RedBisp_Contri} Monopole, dipole and quadrupole of the density term in the reduced bispectrum \eqref{diff_bisp_cont} of a 21cm intensity map, with  {$z_i=1$} and $\ell_1=200$, {normalized {by angular power spectra} as $b_{\ell_1\ell_2\ell_3}/(C_{\ell_1}C_{\ell_2}+C_{\ell_1}C_{\ell_3}+C_{\ell_2}C_{\ell_3})$.}
}
\end{figure}

\begin{figure}
\begin{centering}
\includegraphics[scale=0.44]{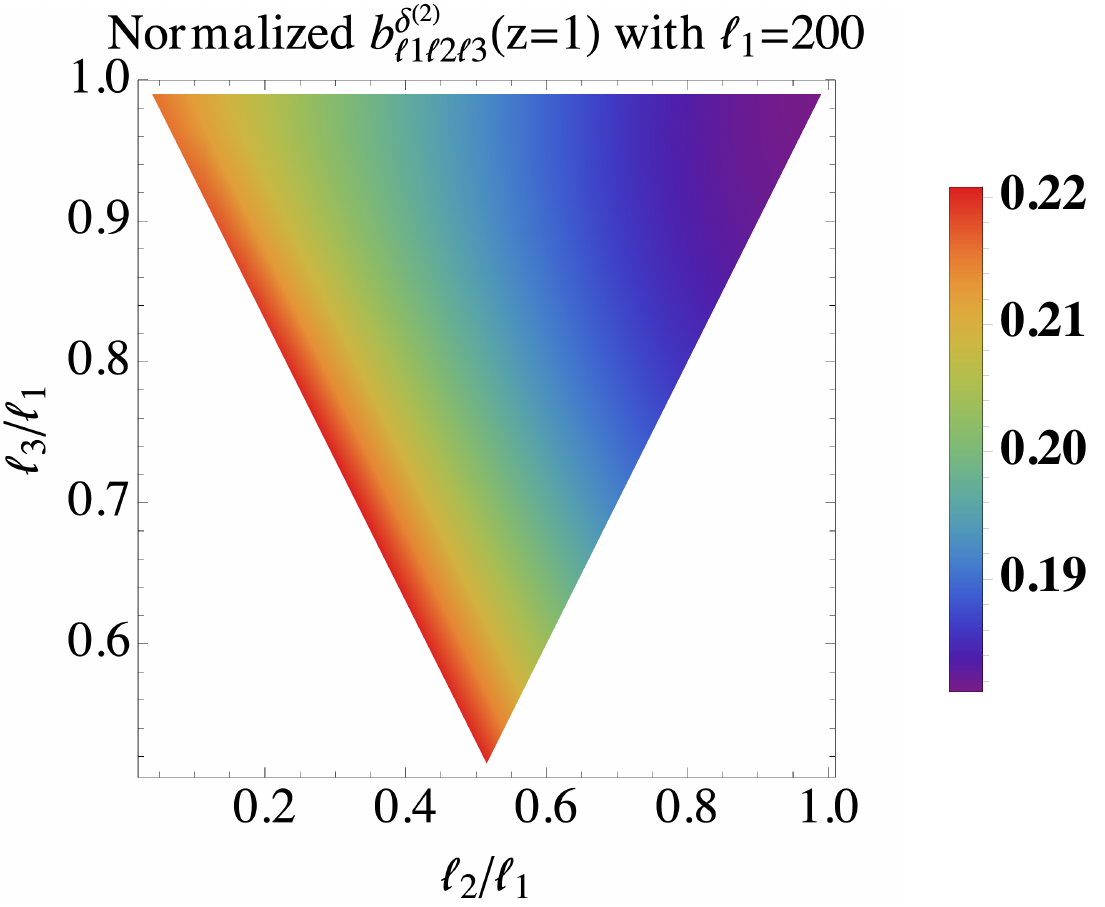} \includegraphics[scale=0.44]{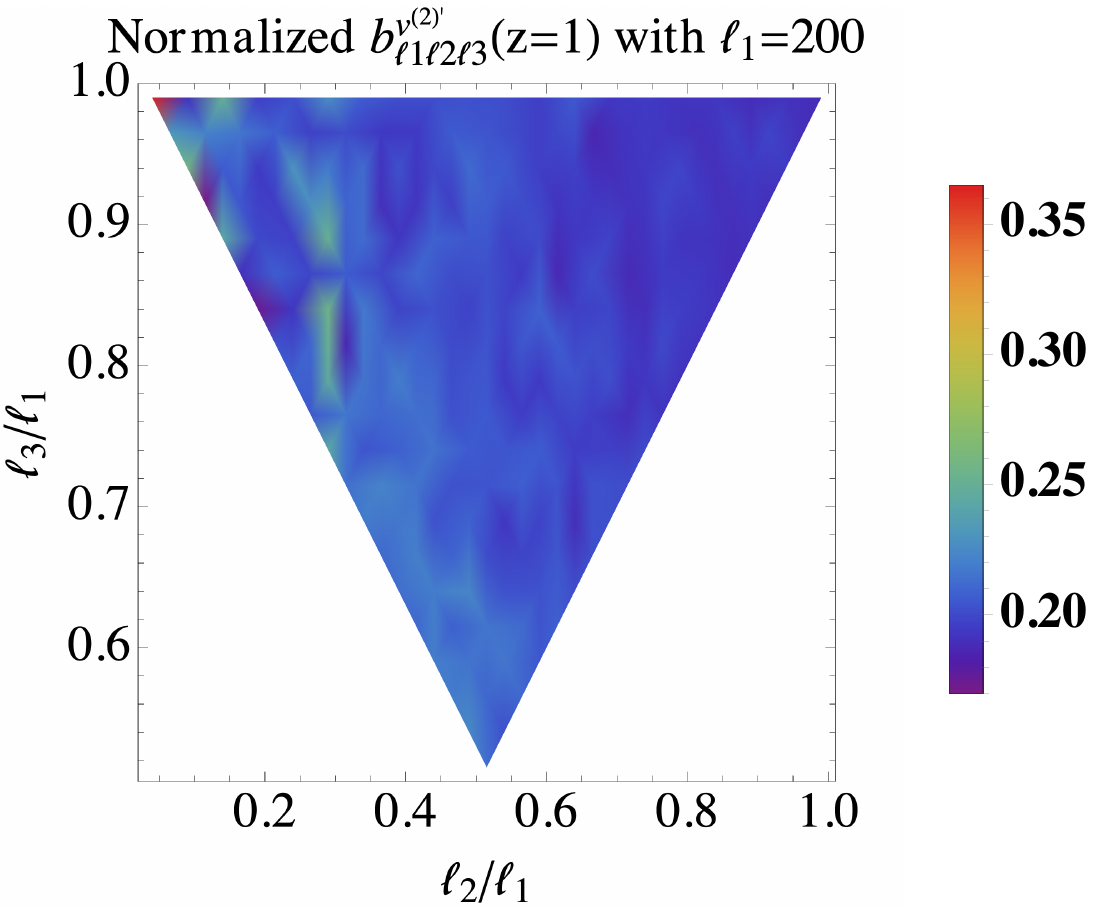}
\includegraphics[scale=0.44]{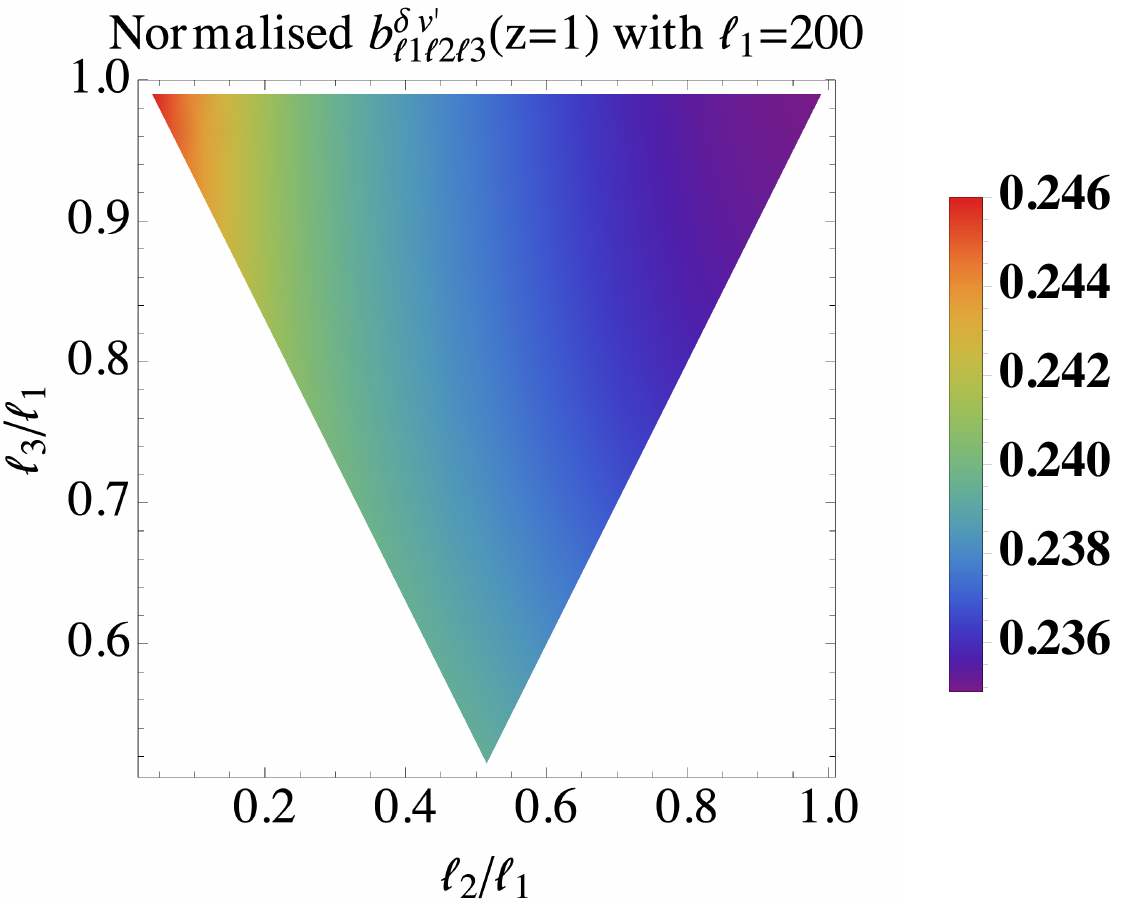}\\
\includegraphics[scale=0.44]{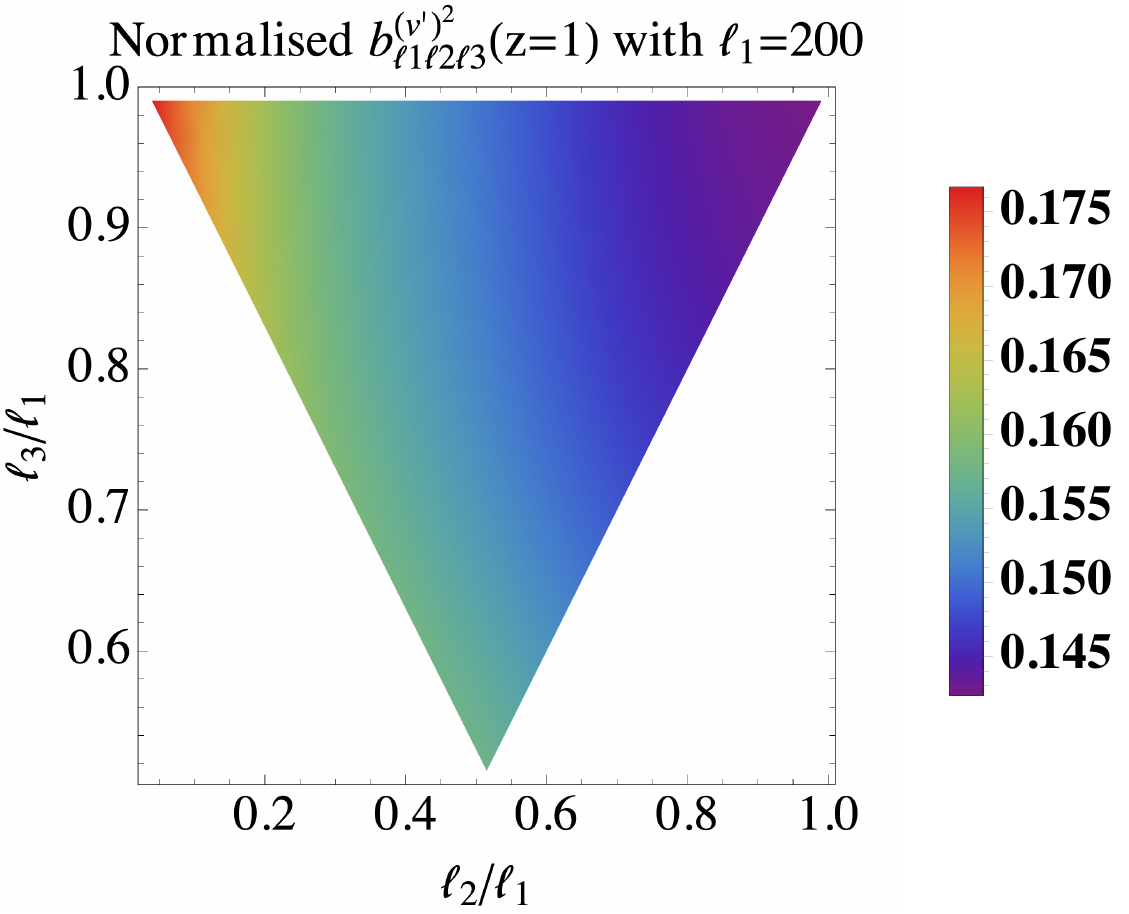}\includegraphics[scale=0.22]{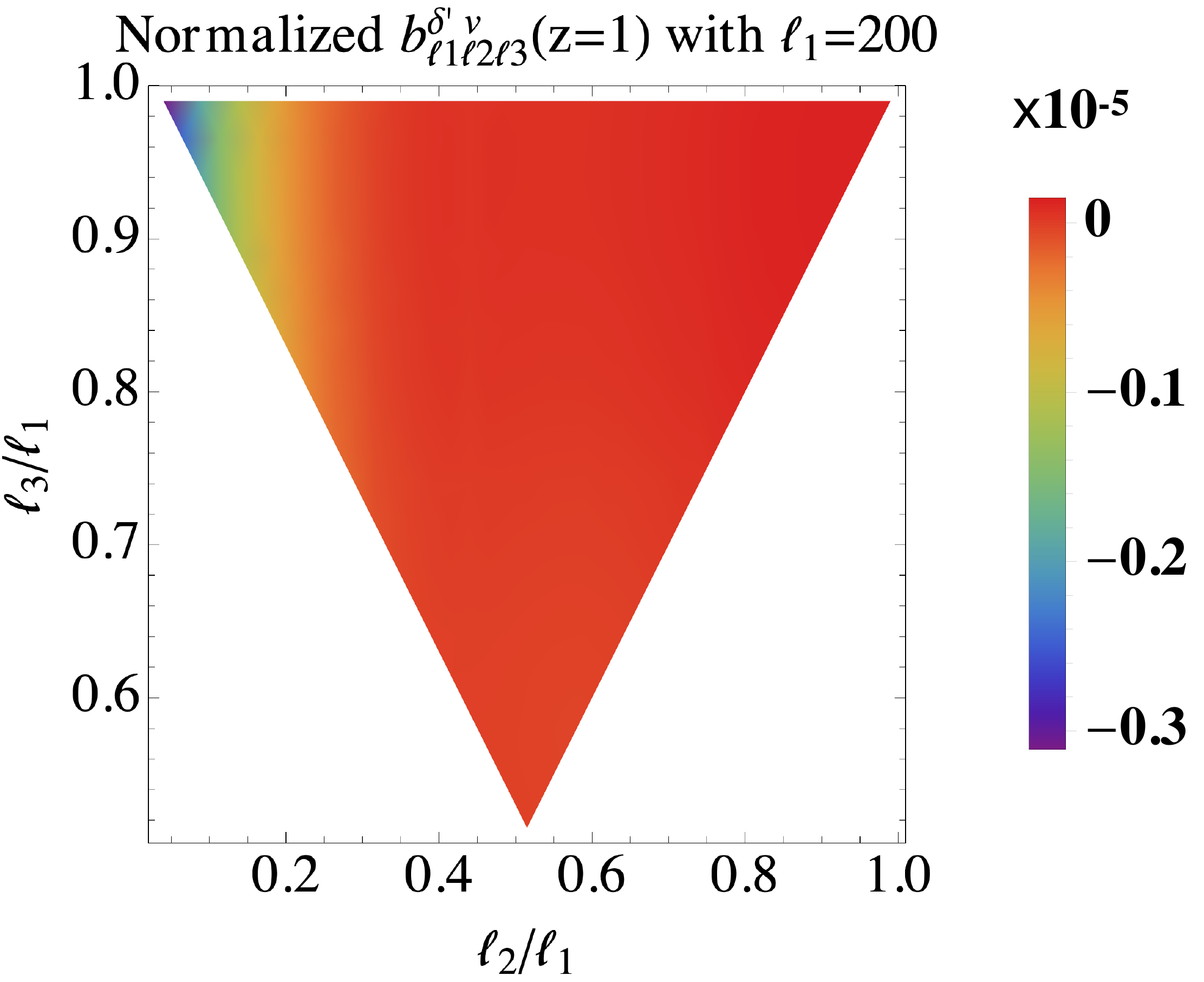}\includegraphics[scale=0.22]{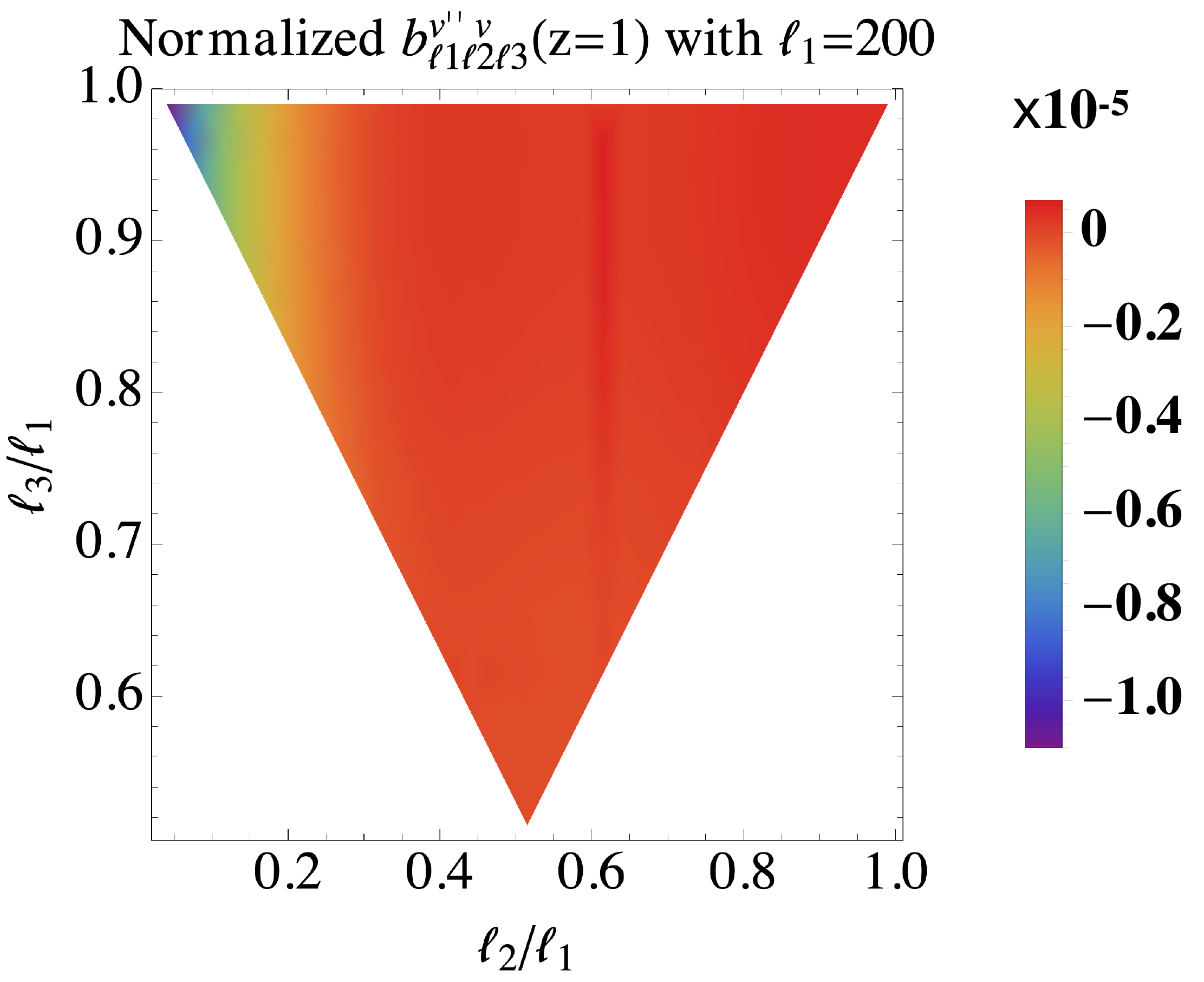}
\par\end{centering}
\caption{\label{fig:All_Bisp}As in Figure \ref{fig:Den_RedBisp_Contri}, for the 6 contributions in \eqref{diff_bisp_cont}, {normalized as above}.
}
\end{figure}

{We use the  \texttt{byspectrum} code \cite{DiDio:2018unb} to compute the angular bispectrum in redshift space. The monopole, dipole and quadrupole of the density contribution}
are shown in contour plots in Figure \ref{fig:Den_RedBisp_Contri}, while Figure \ref{fig:All_Bisp} displays the different contributions in \eqref{diff_bisp_cont}. All plots are normalized relative to the angular power spectrum, as in  \cite{DiDio:2018unb}, and we assume equal redshifts $z_i=1$ and set $\ell_1=200$.

\begin{figure}[ht]
\begin{centering}
\includegraphics[scale=0.27]{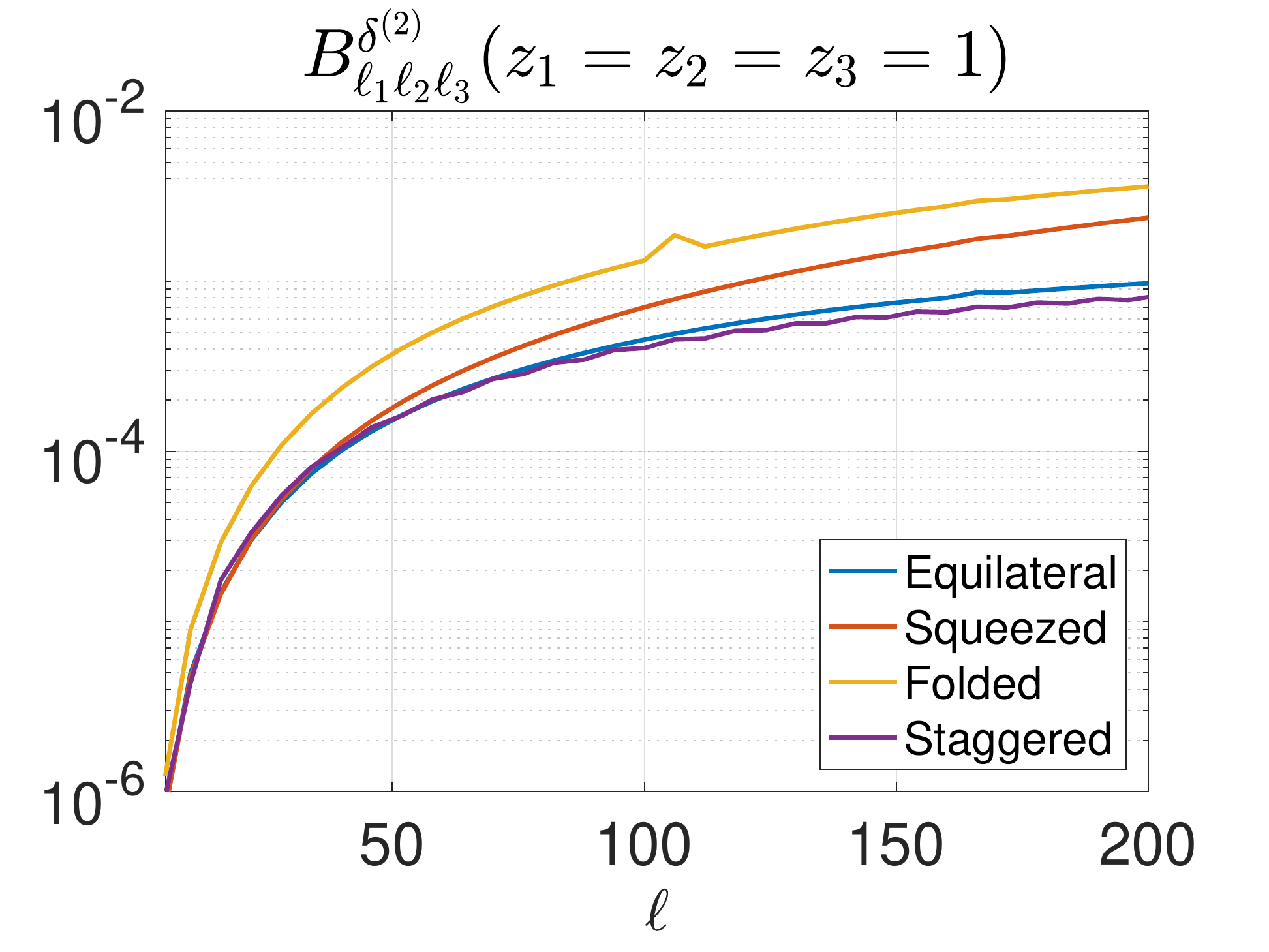}\includegraphics[scale=0.27]{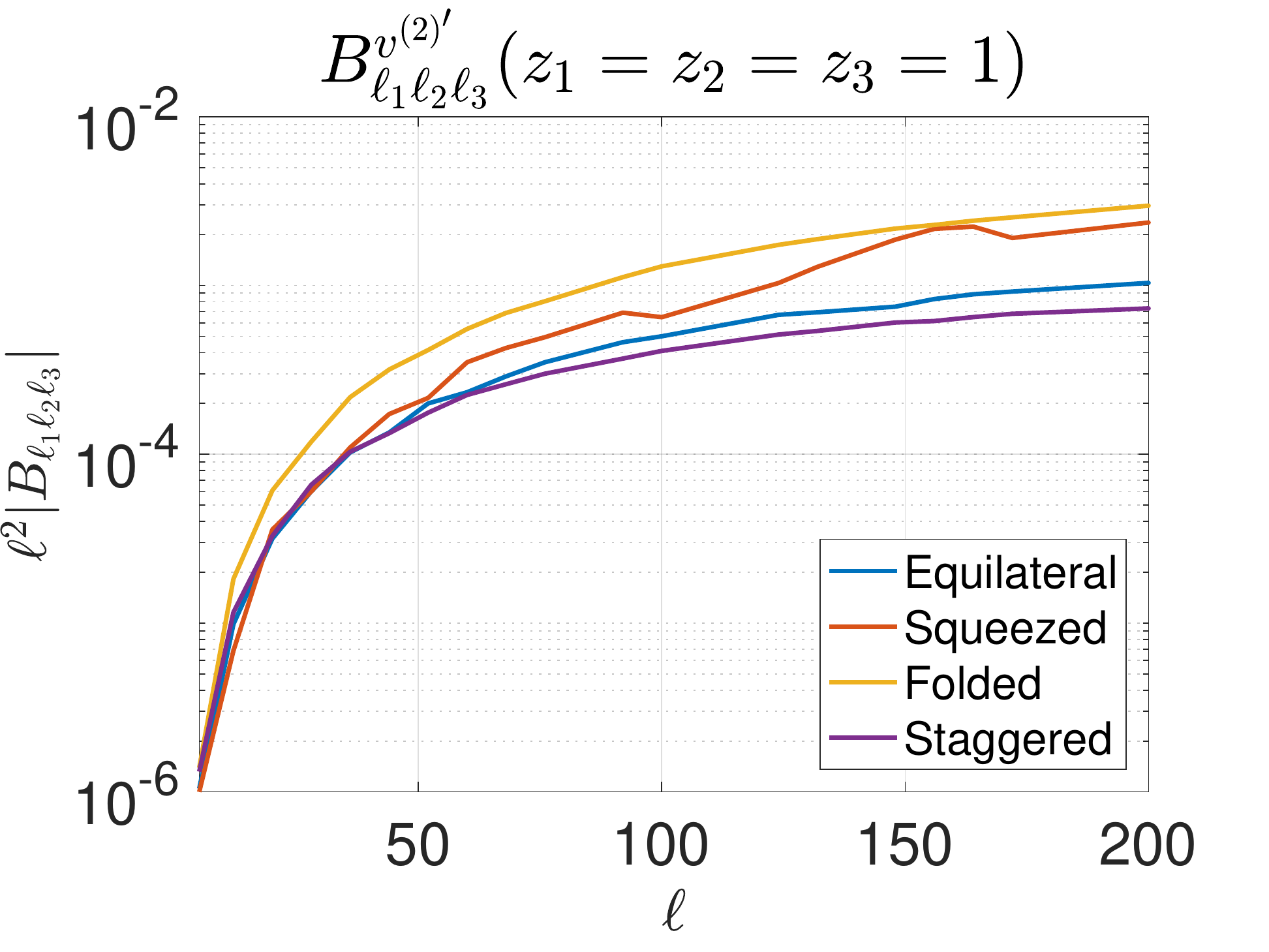}\includegraphics[scale=0.27]{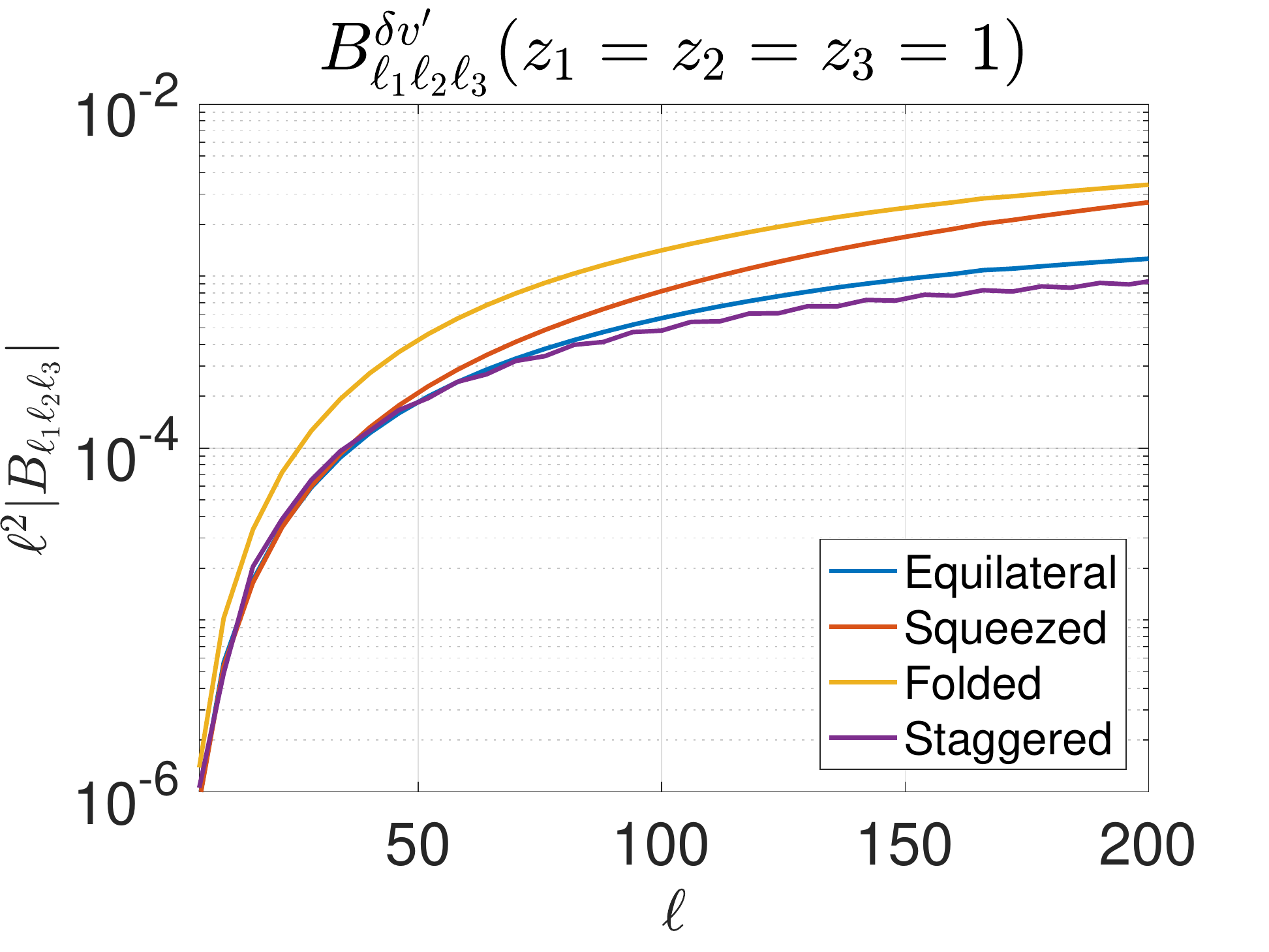} \\\includegraphics[scale=0.27]{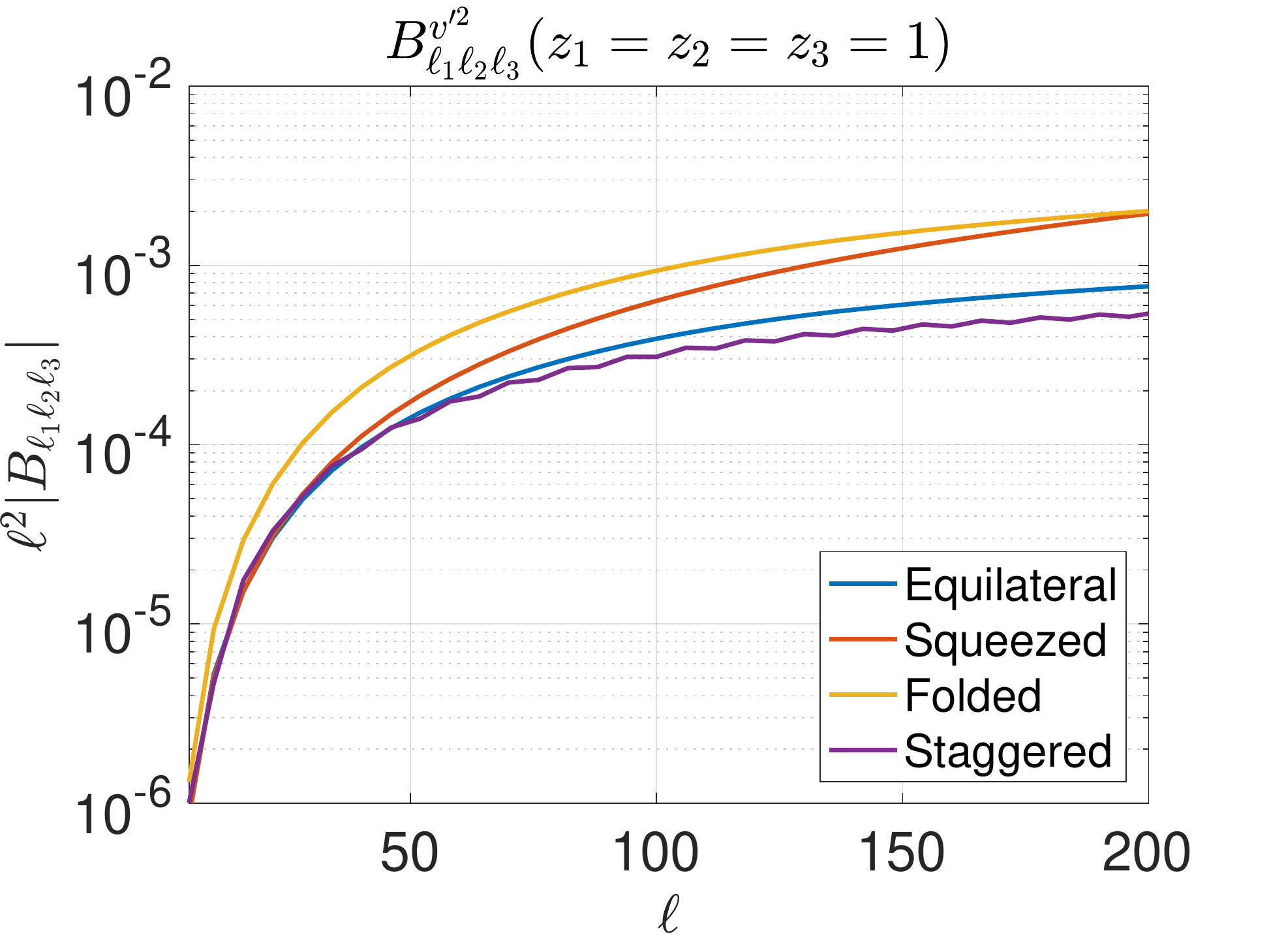}\includegraphics[scale=0.27]{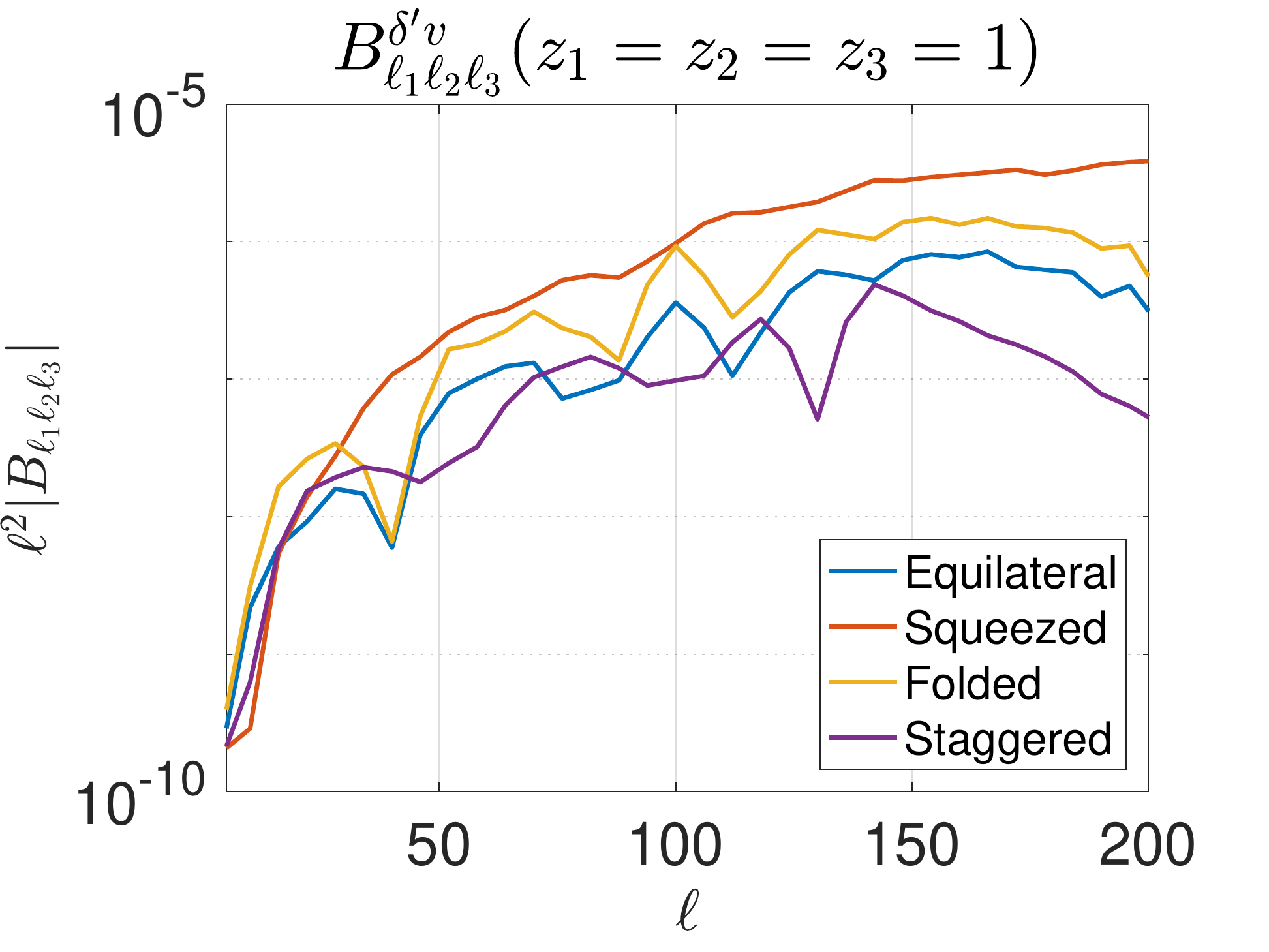}\includegraphics[scale=0.27]{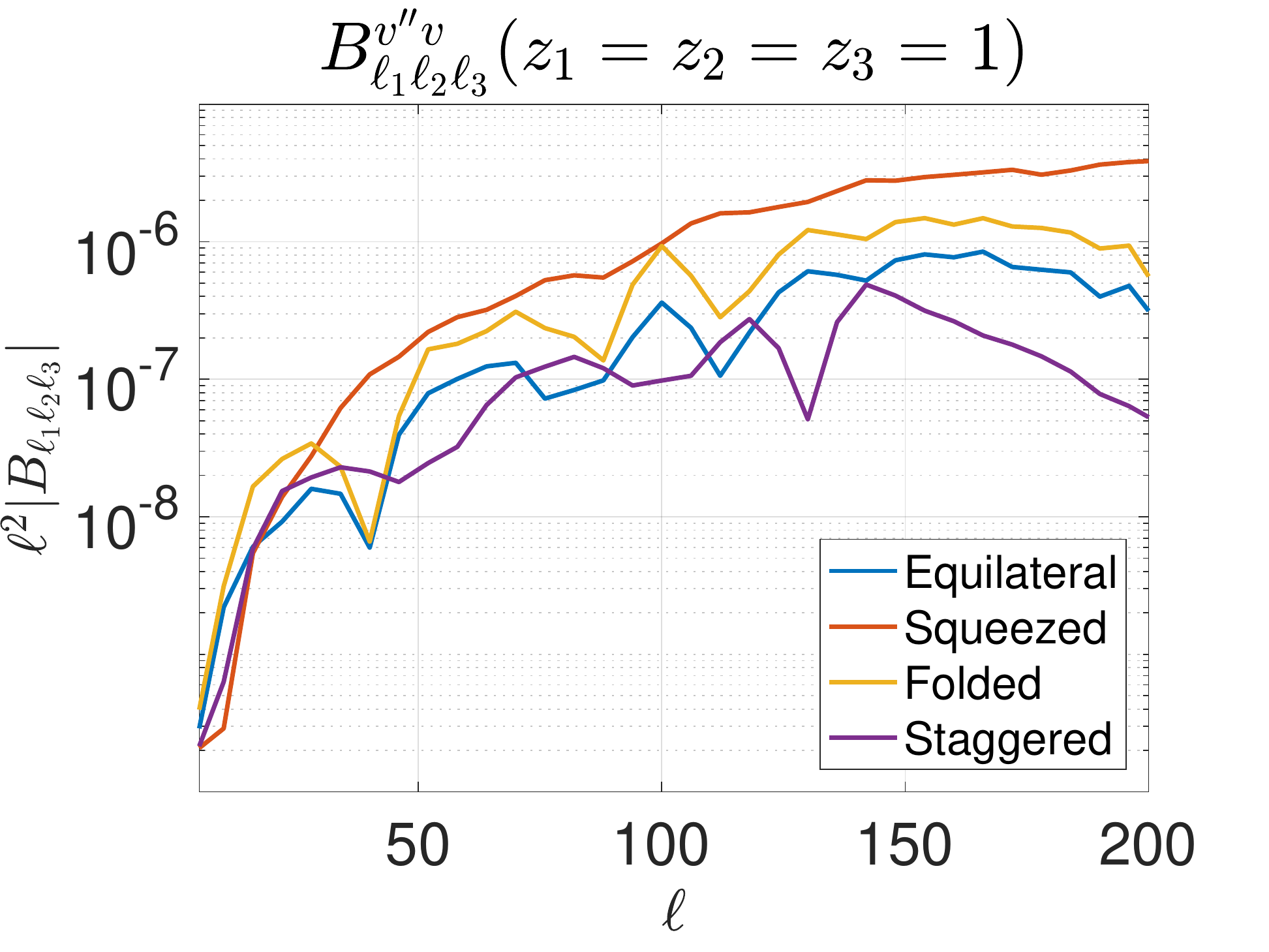}
\par\end{centering}
\caption{\label{fig:RedBisp_Lims}  {Different contributions in  \eqref{diff_bisp_cont} to the
angle-averaged bispectrum \eqref{eq:ang_ave_bis_defi} of a 21cm intensity map, for 3 configurations (with $\ell_1=4$ in the squeezed case) at equal redshifts {$z_i=1$}. In the {lower last two panels} we  used a moving-average filter with window size {2} to smooth numerical features.}
}
\end{figure}

{The  angle-averaged bispectrum is related to the reduced bispectrum as \cite{DiDio:2015bua}
\begin{equation}
B_{\ell_{1} \ell_{2} \ell_{3}}=\sqrt{\frac{\left(2 \ell_{1}+1\right)\left(2 \ell_{2}+1\right)\left(2 \ell_{3}+1\right)}{4 \pi}}\, \left(\begin{array}{lll}
\ell_{1} & \ell_{2} & \ell_{3} \\
0 & 0 & 0
\end{array}\right)\, b_{\ell_{1} \ell_{2} \ell_{3}}\,,
\label{eq:ang_ave_bis_defi}
\end{equation}
where the matrix is the Wigner 3j symbol.
In Figure \ref{fig:RedBisp_Lims}, we compare  the equilateral ($\ell_1=\ell_2=\ell_3$), `squeezed' ($\ell_1=4,\ell_2=\ell_3$) folded  ($\ell_1=\ell_2=\ell_3/2$) and staggered ($\ell_1=\ell$, $\ell_2=[1.5\ell]$, $\ell_3=2\ell$) configurations for the 6 contributions in \eqref{diff_bisp_cont} to the angle-averaged bispectrum. 

Figure \ref{fig:RedBisp_Lims} shows that the equilateral shape makes the smallest  contribution to the total angle-averaged bispectrum. {For all three shapes, the {nonlinear RSD terms} $B^{\delta'v}$ and $B^{v''v}$ are negligible. {For the other three nonlinear RSD terms, the reduced bispectra $b^{v^{(2)'}}, b^{\delta v'}$ and $b^{v^{\prime 2}}$} are all comparable to the density term $b^{\delta^{(2)}}$. Furthermore, they are all positive.\footnote{{Note that the angle-averaged bispectra can give an opposite sign to the corresponding reduced bispectra since Wigner 3j symbols are negative when $m_i=0$ and $(\ell_1+\ell_2+\ell_3)/2$ is odd.}} This can be seen since there is no sign change in the plots of Figure \ref{fig:RedBisp_Lims}, while in Figure~\ref{fig:All_Bisp}, we see that these terms are all positive  for $\ell_1=200$. In addition, the four dominant contributions have approximately equal magnitudes for the three shapes. This {anticipates what we find below:} that including RSD increases the bispectrum by  a {significant} factor.}}  

\begin{figure}
\begin{centering}
\includegraphics[scale=0.38]{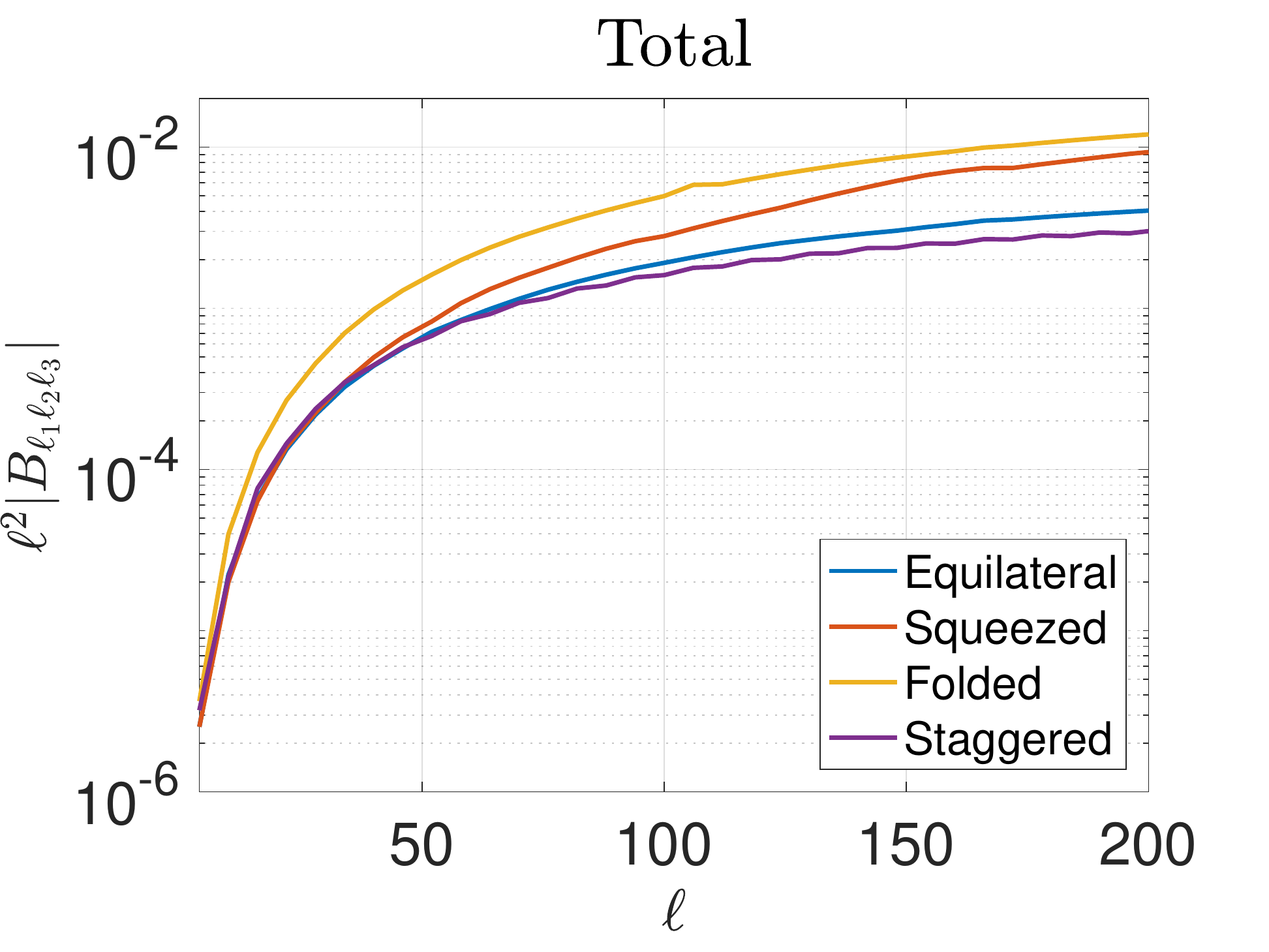}
\includegraphics[scale=0.38]{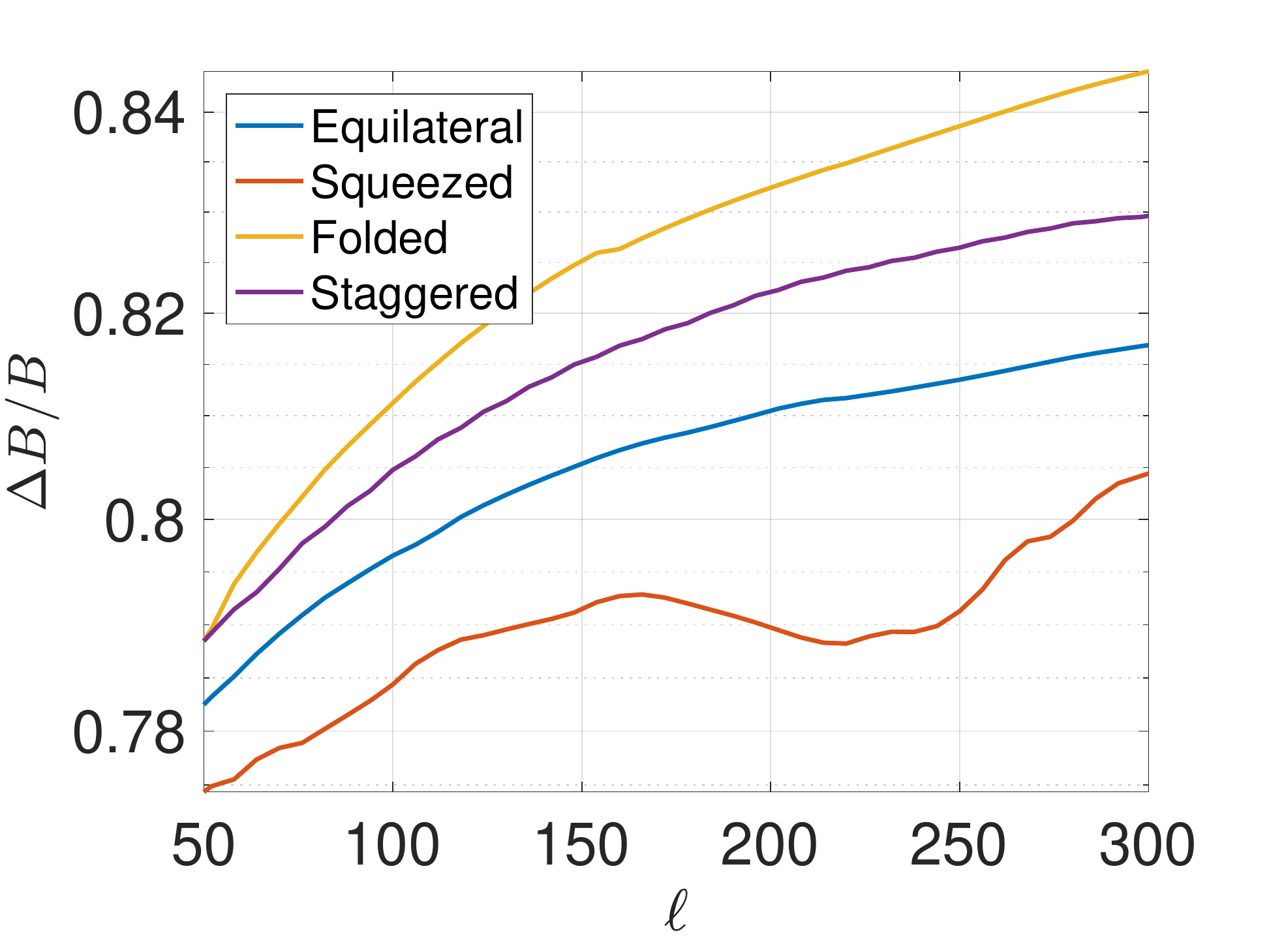}
\par\end{centering}
\caption{\label{fig:tot_bisp} {\em Left:} Total angle-averaged bispectrum from Figure \ref{fig:RedBisp_Lims}. {\em Right:} RSD fractional contribution \eqref{dbb1} to this angle-averaged bispectrum.}
\end{figure}

{In order to quantify the RSD contribution  to the 21cm intensity bispectrum, we show the total angle-averaged bispectrum in Figure \ref{fig:tot_bisp} ({left} panel) and the fractional RSD contribution to the total angle-averaged bispectrum, $\Delta B/B$ in} Figure \ref{fig:tot_bisp} ({right} panel), where we define
\begin{equation}\label{dbb1}
    \frac{\Delta B}{B}= \frac{B_{\ell_1\ell_2\ell_3}[\mbox{with RSD}] -
    B_{\ell_1\ell_2\ell_3}[\mbox{no RSD}]}{ B_{\ell_1\ell_2\ell_3}[\mbox{with RSD}]}\,.
\end{equation}
Here `no RSD' denotes the bispectrum with only the $b^{\delta^{(2)}}$ contribution in \eqref{diff_bisp_cont} (corresponding to the first 3 terms in \eqref{eq:delta_sec_ord}) and with no RSD in the linear term $\Delta^{(1)}$, i.e., with $\Delta^{(1)}=\delta^{(1)}$.

{Interestingly, RSD make up 80\% or more of the total signal for all three triangular shapes considered here, so that RSD increases the bispectrum by a factor $\sim$5.}
{Figure \ref{fig:tot_bisp} corresponds to the maximal RSD contribution, since it is based on zero-width redshift bins. Nevertheless, we expect the RSD contribution to remain dominant for finite but thin redshift bins, similar to the case of the angular power spectrum \cite{Fonseca:2019qek}.}

\section{Detectability of the bispectrum} \label{sec:SNR_Calcu}

{In order to assess the detectability of the 21cm intensity bispectrum, we consider
two next-generation intensity mapping surveys, with SKA-MID \cite{Bacon:2018dui} and HIRAX \cite{Newburgh:2016mwi}. The survey specifications are shown in Table \ref{tab:ExpSpeci}.
21cm intensity surveys can be performed in {(a)} single-dish (SD) mode, where the individual auto-correlations from each dish are simply summed, or  in {(b)} interferometric (IF) mode, where the cross-correlations of all dishes are combined. Surveying in SD mode captures larger scales, while IF mode surveys can resolve smaller scales (see e.g. \cite{Bull:2014rha,Alonso:2017dgh}). 
SKA-MID is better adapted to SD mode, while HIRAX is designed for IF mode.}

\begin{table}[tbp]
\centering
\begin{tabular}{c c c c c c c c c}
\hline
\hline
\noalign{\vskip 0.1cm}
Survey & $f_{\mathrm{sky}}$ & $N_{\mathrm{d}}$ & $t_{\mathrm{tot}}$({hr}) & $T_{\mathrm{ins}}$(K) &  $D_{\mathrm{d}}$(m) & {redshift}\\
\noalign{\vskip 0.1cm}
\hline
\hline
\noalign{\vskip 0.1cm}
SKA & {0.48} & {197} & {10,000} & 28  & 15 & {0.3--3} \\
HIRAX & 0.36 & 1024 & {10,000} & 50  & 6 & {0.8--2} \\
\hline
\hline
\end{tabular}
\caption{\label{tab:ExpSpeci} Specifications that we assume for HIRAX and SKA. 
}
\end{table}

{Before we consider the noise and the signal for these 21cm intensity surveys, we need to determine the appropriate limiting angular scales.}
The $\ell_i$ in the bispectrum should each satisfy the condition 
\be
\ell_{\mathrm{min}}\leq \ell_i \leq\ell_{\mathrm{max}}\,,
\ee 
where the lower limit depends on the survey and on foreground cleaning, while the upper limit depends on the survey and the modelling of nonlinearity.  
{Foreground cleaning  removes large-scale radial  modes in Fourier space, with $k_\|\lesssim 0.01h/$Mpc, although this limit can be lowered by the technique of reconstructing large-scale modes using short-scale measurements \cite{Zhu:2016esh,Modi:2019hnu}. In angular harmonic space, modes with $\ell \lesssim 5$ in the power spectrum are effectively lost \cite{Witzemann:2018cdx,Fonseca:2019qek,Schmit:2018rtf}.} We therefore impose
\be \label{ellmin}
{\ell_{\mathrm{min}}^{\rm \,fground} \approx 5}\,.
\ee
{The survey sky area $\Omega_{\mathrm{sky}}= 4\pi f_{\mathrm{sky}}$ determines the largest possible scale included, {which imposes the theoretical lower limit} \cite{Fonseca:2015laa}:
\begin{equation}
    {\ell_{\mathrm{min}}^{\rm \,sky}}=1+ {\rm int}\Big(\frac{\pi}{\sqrt{\Omega_{\rm sky}}}\,\Big)\,,
\end{equation}
where `int' denotes the integer part. For SKA and HIRAX, this is below 5, and hence \eqref{ellmin} applies. In fact, IF mode may have $\ell_{\mathrm{min}}$ much larger than 5, since {it is} the minimum baseline of an interferometer {that} sets the largest observable mode \cite{Schmit:2018rtf}. Equivalently, there is a maximum scale determined by the field of view in IF pointings \cite{Bull:2014rha}. The field of view is determined by the effective beam:
\be \label{eq:thetab}
\theta_{\mathrm{b}}=1.22\,\frac{\lambda_{21}}{D_{\mathrm{d}}}\,(1+z)\,,
\ee
where $\lambda_{21}$ is the rest-frame 21cm wavelength and $D_{\rm d}$ is the dish diameter (see Table \ref{tab:ExpSpeci}). Then
\bea
    {\ell_\mathrm{min}^{\rm \,IF}}(z) &\approx & \frac{2\pi}{\theta_\mathrm{b}(z)}
    \\
& \approx&  \frac{147}{1+z}\quad \mbox{for HIRAX}\,.
\eea} 

For the angular power spectrum with equal redshift correlations,  a theoretical maximum $\ell$ condition is imposed by the range of validity of  the tree-level angular power spectrum \cite{Fonseca:2019qek}:
\begin{equation}\label{lfgmin}
{\ell_{\rm max}^{\rm \,nl}}(z)= r(z)\,k_{\rm nl}(z)\,,\quad k_{\rm nl}(z)=k_{\rm nl}(0)(1+z)^{2/3}~{\rm Mpc}^{-1}\,.
\end{equation}
For $C_\ell$, the nonlinear scale is typically taken as $k_{\rm nl}(0)=0.2h/$\,Mpc  (see also \cite{DiDio:2013sea}). In order to reflect the greater sensitivity of the bispectrum to nonlinearity, we
follow \cite{Maartens:2019yhx} and assume $k_{\rm nl}(0)=0.1h/$\,Mpc, i.e. half of the value for the power spectrum (see also \cite{Yankelevich:2018uaz}). 
\begin{figure}[ht]
\begin{centering}
\includegraphics[scale=0.5]{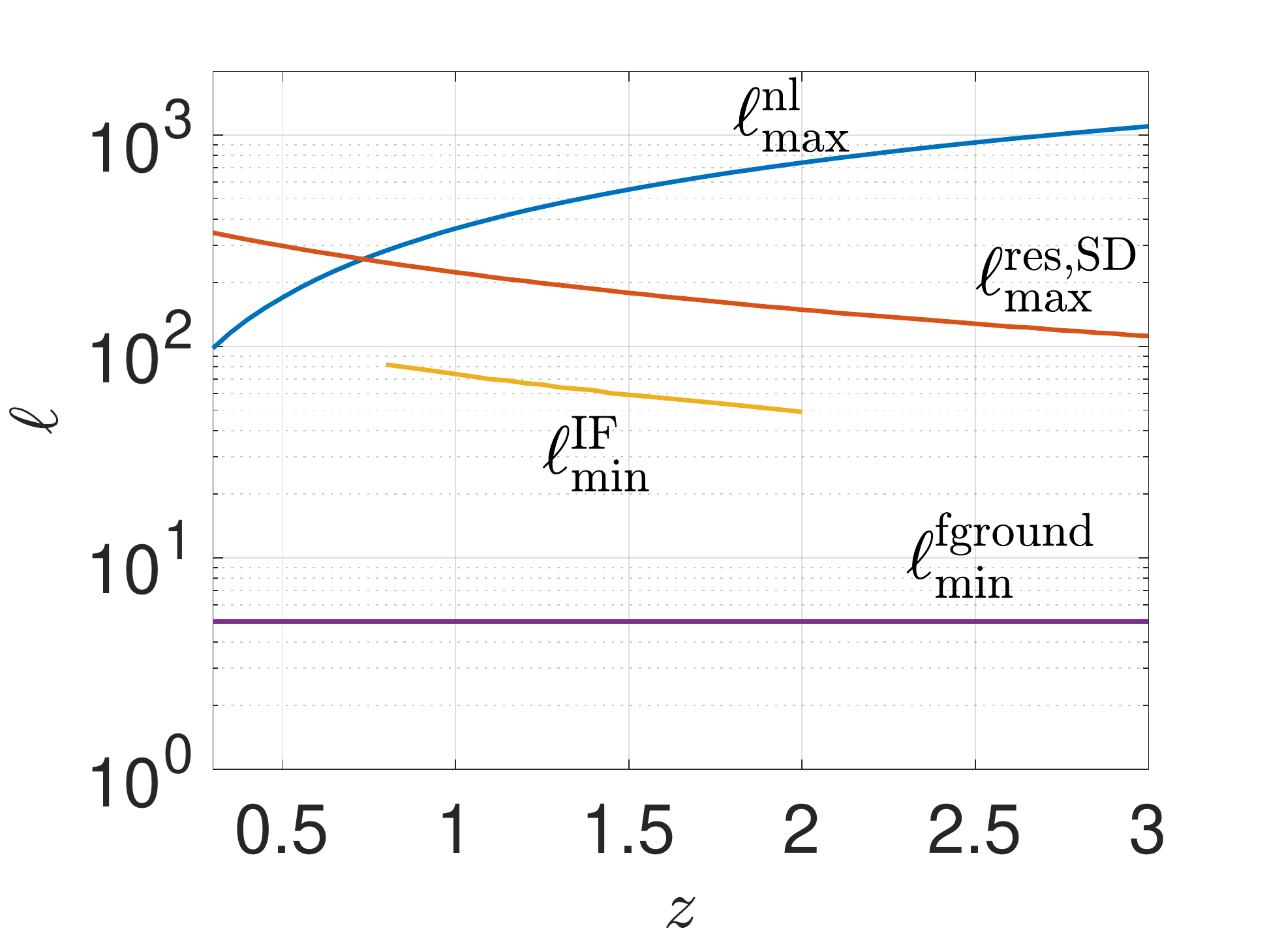}
\par\end{centering}
\caption{\label{fig:LimScales} Minimum and maximum  scales for SKA (SD) and HIRAX (IF) surveys. }
\end{figure}

{There is also an experimental maximum imposed by the angular resolution of the array: $k_{\perp,{\rm max}}\approx 2\pi D_{\rm res}/(r\lambda)$  where $D_{\rm res}$ is the diameter of the receiving area of the beam array and $\lambda=\lambda_{21}(1+z)$. Thus
\be \label{ellm2}
\ell_{\rm max}^{\rm \,res}(z)\approx {2\pi D_{\rm res} \over \lambda_{21}(1+z)}\,.
\ee
In SD mode, $D_{\rm res}=D_{\rm d}$ and then,
\be \label{ellm3}
\ell_{\rm max}^{\rm \,res,SD}(z)\approx {2\pi D_{\rm d} \over \lambda_{21}(1+z)}\approx {449 \over 1+z} \quad \mbox{for SKA} \,.
\ee
This can be smaller than \eqref{lfgmin} at high $z$. In IF mode, $D_{\rm res}$ is the maximum baseline, which  is $\approx271\,$m  for HIRAX, so that 
\be \label{ellm4}
\ell_{\rm max}^{\rm \,res,IF}(z)\approx {2\pi D_{\rm max} \over \lambda_{21}(1+z)}\approx {8108 \over 1+z} \quad \mbox{for HIRAX} \,.
\ee
For all $z$, \eqref{ellm4} is much larger than  \eqref{lfgmin}.} 

{In summary, for the surveys considered, we have 
\bea
\mbox{SKA (SD):} &~~& 
\ell_{\rm min}= \ell_{\mathrm{min}}^{\rm \,fground}= 5\,,\qquad \qquad~
\ell_{\mathrm{max}}= {\rm min} \big\{\ell_{\mathrm{max}}^{\rm \,nl}(z)\,, \, 449 (1+z)^{-1} \big\},~~~~
\\
\mbox{HIRAX (IF):} &~~&
\ell_{\rm min}= \ell_{\mathrm{min}}^{\rm \,IF}= 147 (1+z)^{-1}\,,~~ 
\ell_{\mathrm{max}}= \ell_{\mathrm{max}}^{\rm \,nl}(z) \,.
\eea
The different limiting scales are shown in Figure \ref{fig:LimScales}.}

\subsection{{Intensity mapping noise}}

{To determine the detectability of the angular bispectrum for 21cm intensity surveys with SKA and HIRAX, we have to study the noise for each survey. The 21cm noise power spectrum is dominated by thermal noise (from the sky and the instrument), shot noise can be neglected  \cite{Fonseca:2019qek,Chang:2007xk}. For SD and IF modes, the dimensionless noise power spectrum in a single redshift bin has the form \cite{Bull:2014rha,Alonso:2017dgh}
\bea \label{genn}
{\mathcal{N}_{\ell}}(z)=\frac{2\pi f_{\mathrm{sky}}}{t_{\mathrm{tot}}\, \Delta\nu(z)}\left[\frac{T_{\mathrm{sys}}(z)}{\bar{T}_{\mathrm{HI}}(z)}\right]^{2}\,{\alpha_\ell(z) \over \beta_\ell(z)^2} \,.
\eea
Here $t_{\mathrm{tot}}$ is the observing time (given in Table \ref{tab:ExpSpeci}) and $\Delta \nu$ is the bandwidth of the redshift bin with width $\Delta z$: 
\begin{equation}
\label{eq:Dnu}
\Delta\nu=\nu_{21}\,\frac{\Delta z}{(1+z)^{2}}\,.
\end{equation}
The system temperature is made up of instrument and sky contributions:
\begin{equation}
T_{\mathrm{sys}}(z)=T_{\mathrm{ins}}+60\left[\frac{300(1+z)}{\nu_{21} / {\rm MHz}}\right]^{2.55}~ \mathrm{K}\,,\label{eq:T_Sys}
\end{equation}
where $T_{\mathrm{ins}}$ is the instrument temperature (see Table \ref{tab:ExpSpeci}) and $\nu_{21}=1420\,$MHz. The sky temperature is an approximate fit to observations and other fits can be used. The background 21cm brightness temperature is given by \cite{Villaescusa-Navarro:2018vsg}
\be
\bar{T}_{\mathrm{HI}}(z)= 189h
\, \frac{H_0(1+z)^2}{H(z)}\, \Omega_{\mathrm{HI}}(z)~~{\rm mK}.
\ee
Given the paucity of observations, $\Omega_{\rm HI}(z)= \bar{\rho}_{\mathrm{HI}}(z)/\rho_{\rm crit}(0)$ is not well constrained and different simulations can lead to significantly different results.
We use the fit \cite{Fonseca:2019qek}
\begin{equation}
    \bar{T}_{\mathrm{HI}}(z)=0.056+0.232\,z-0.024\,z^{2}~~ \mathrm{mK}\,.\label{eq:T_Bar_HI_SKA}
\end{equation}

Finally, the dish density factor $\alpha_\ell$ and effective beam $\beta_\ell$ differ for SD and IF modes as follows \cite{Bull:2014rha,Fonseca:2019qek,Jalilvand:2019bhk,Karagiannis:2019jjx}.
\begin{itemize}
    \item 
\underline{SD mode}:
\bea 
\alpha^{\rm SD}_\ell(z) &=&{1\over N_{\rm d}}\,,
\\
\beta^{\rm SD}_\ell(z) &=&\exp\left[-\frac{\ell(\ell+1)}{16\ln 2}\,\theta_{{\mathrm{b}}}(z)^2 \right]\,, \label{eq:theta_beam}
\eea
where $N_{\rm d}$ is the number of dishes (see Table \ref{tab:ExpSpeci}) and $\theta_\mathrm{b}$ is given by \eqref{eq:thetab}.

\item
\underline{IF mode}:
\bea 
\alpha^{\rm IF}_\ell(z) &=& \left[{\lambda(z)^2\over A_{\rm eff}}\right]^2\,{1 \over n_{\rm b}({z},\ell)}\,,\label{ifn1}
\\ \label{ifn2}
\beta^{\rm IF}_\ell(z) &=& {\theta_{\mathrm{b}}(z)},
\eea
where $A_{\mathrm{eff}}=0.7\pi D_{\mathrm{d}}^2/4$ is the effective dish area and $n_{\rm b}$ is {the baseline density in the image plane,} determined by the dish distribution. 
The forms \eqref{ifn1}, \eqref{ifn2} apply in the case where the pointings (which cover the field of view) are done sequentially. 

HIRAX is an example of a square-packed array. Following \cite{Karagiannis:2019jjx},  we use the fitting formula from \cite{Ansari:2018ury}:
\bea
n_{\rm b}({z},\ell) &=&{N_{\rm d}}(1+z)^2\, \left({\lambda_{21} \over D_{\rm d}}\right)^2 \left[ {a_1+a_2({L/L_{\rm s}}) \over 1+a_3({L/L_{\rm s}})^{a_4}}  \right]\,\exp\left[-\left({L\over L_{\rm s}} \right)^{a_5} \right]\,,\label{nb1}\\
{L(z,\ell)} &=& {\lambda_{21}\over 2\pi}\,(1+z)\,\ell\,.
\label{nb2}
\eea
Here $L$ is the baseline radial length, ${L_{\rm s}}=D_{\rm d} \sqrt{N_{\rm d}}= {192}\,$m is the length of the square side and the fitting parameters are
$a_I=\big(0.4847,-0.3300,1.3156,1.5974,6.8390\big)$. See Figure \ref{nbh}.
\begin{figure}[!ht]
\begin{centering}
\includegraphics[scale=0.38]{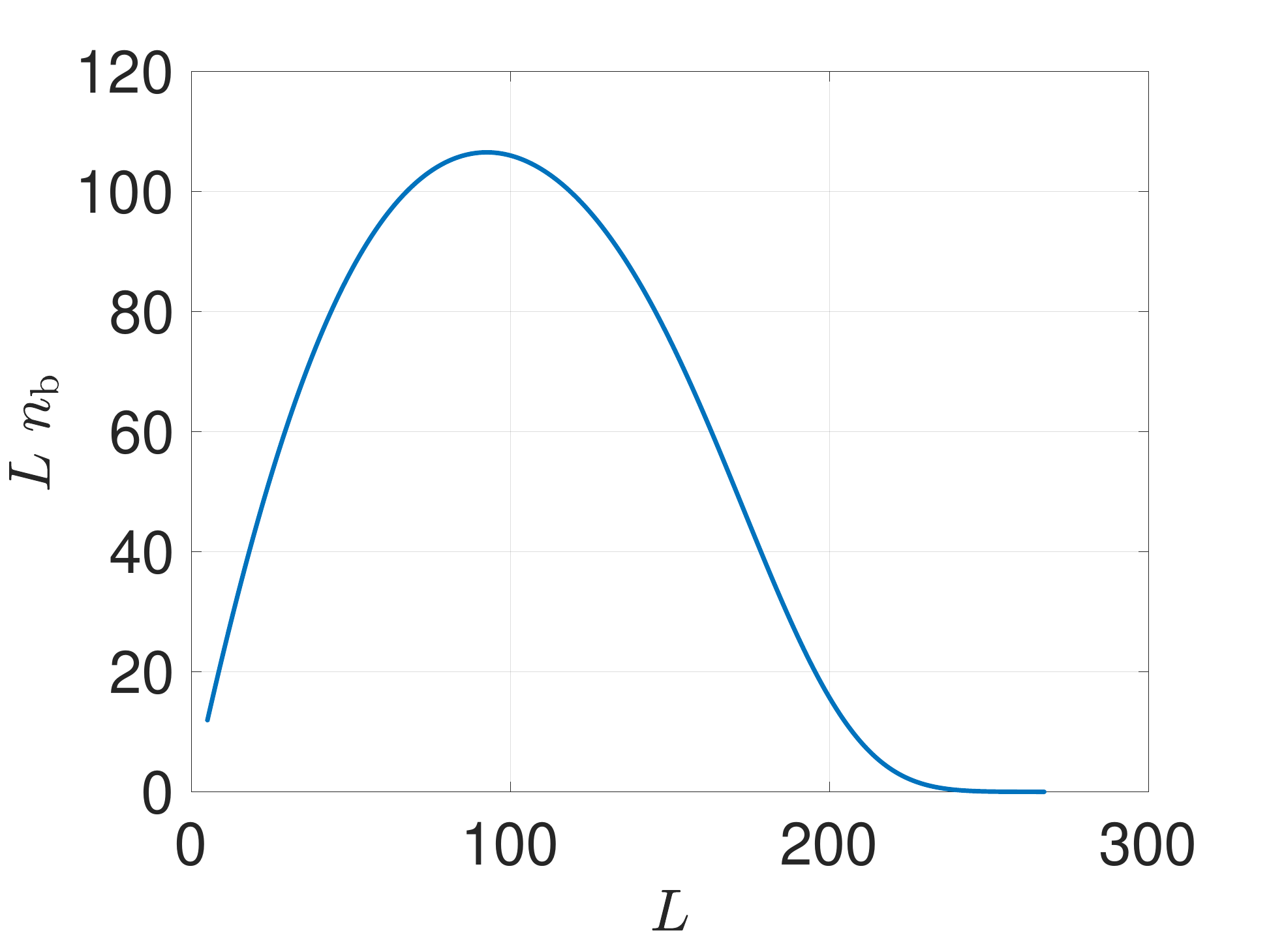}\includegraphics[scale=0.38]{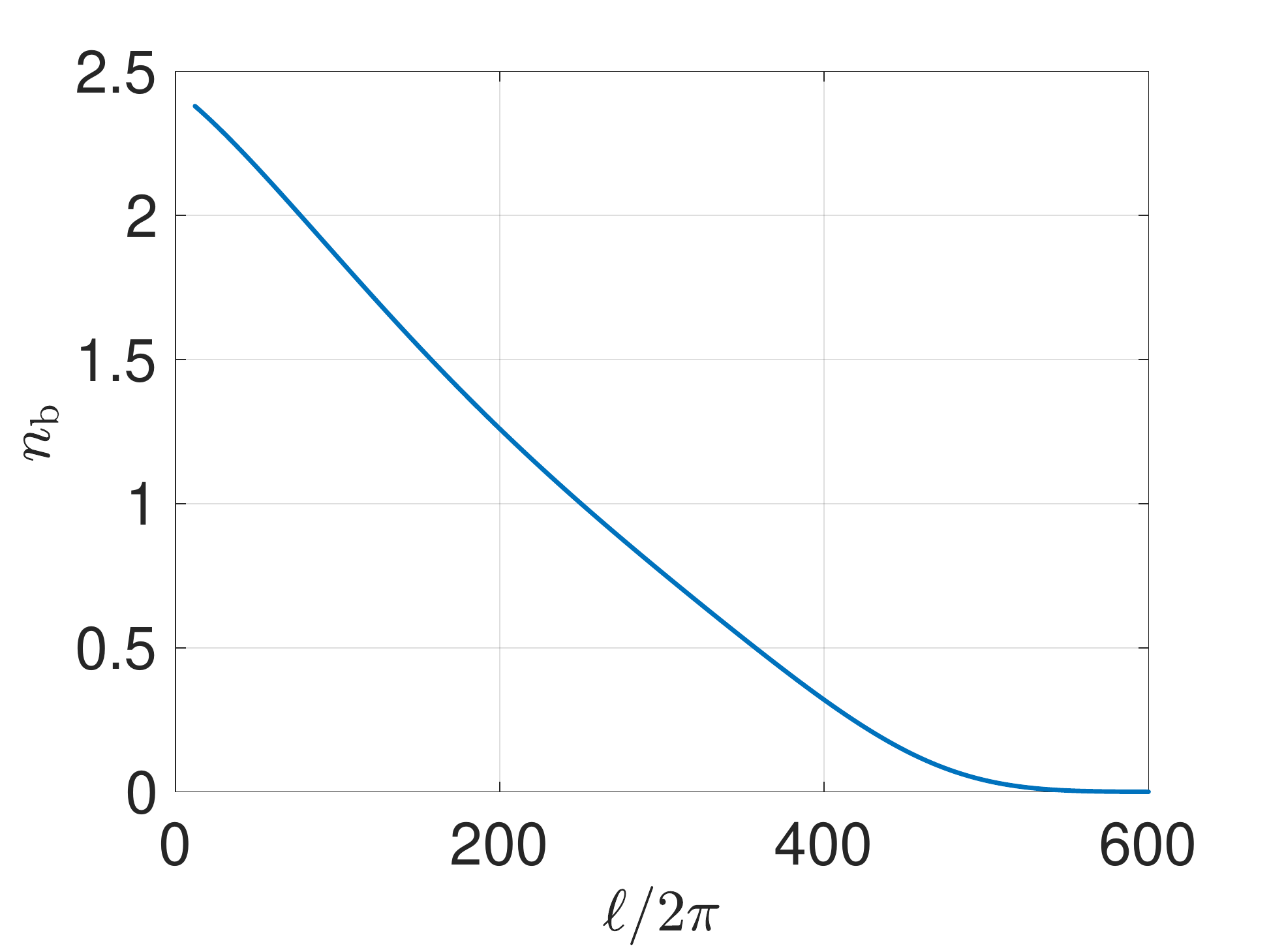}
\par\end{centering}
\caption{\label{nbh} 
Baseline density for HIRAX at {$z=1$} {as a function of $L$ (in m) and $\ell$.}}
\end{figure}
\end{itemize}}

\begin{figure}
\begin{centering}
\includegraphics[scale=0.38]{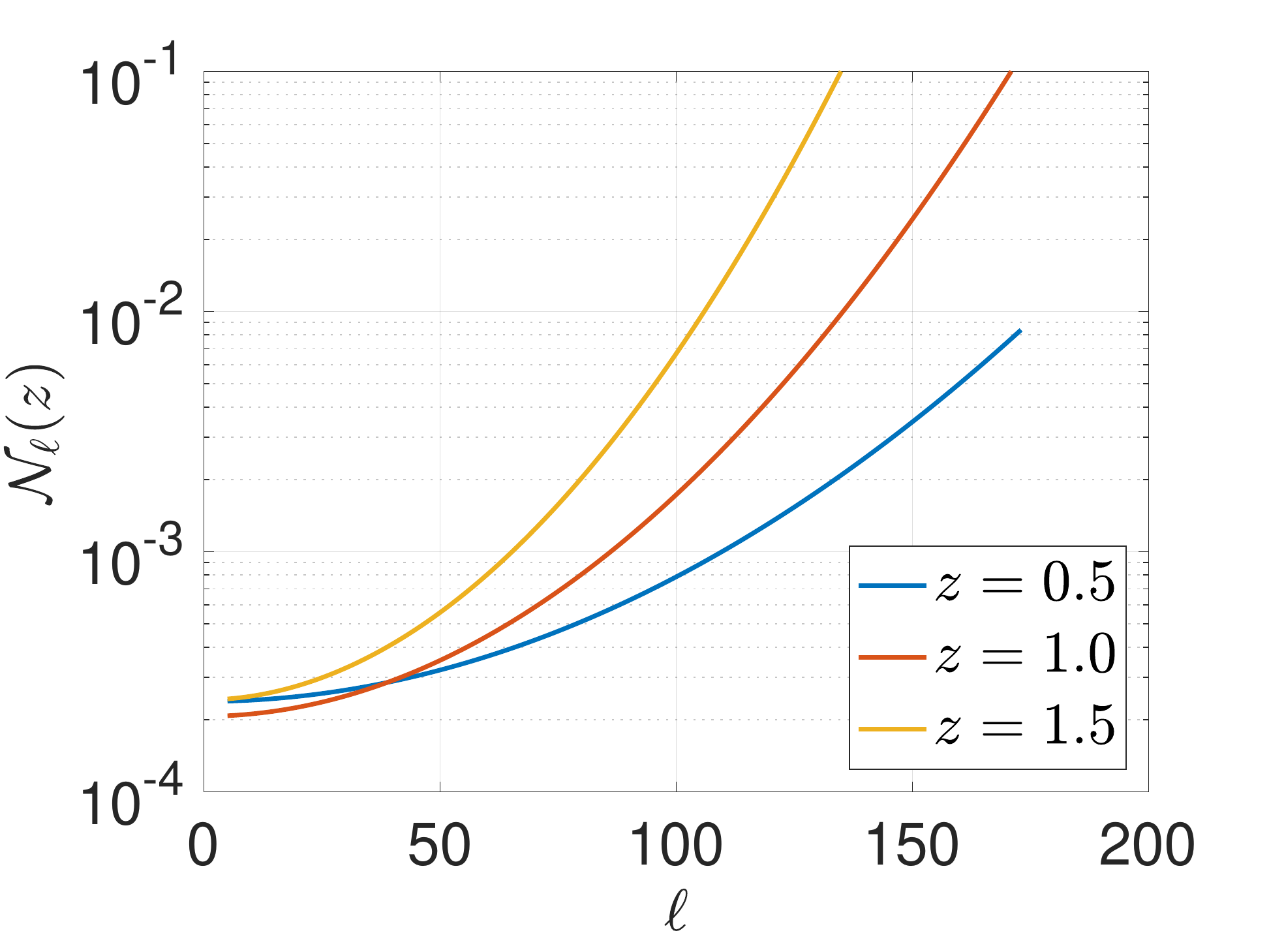}
\includegraphics[scale=0.38]{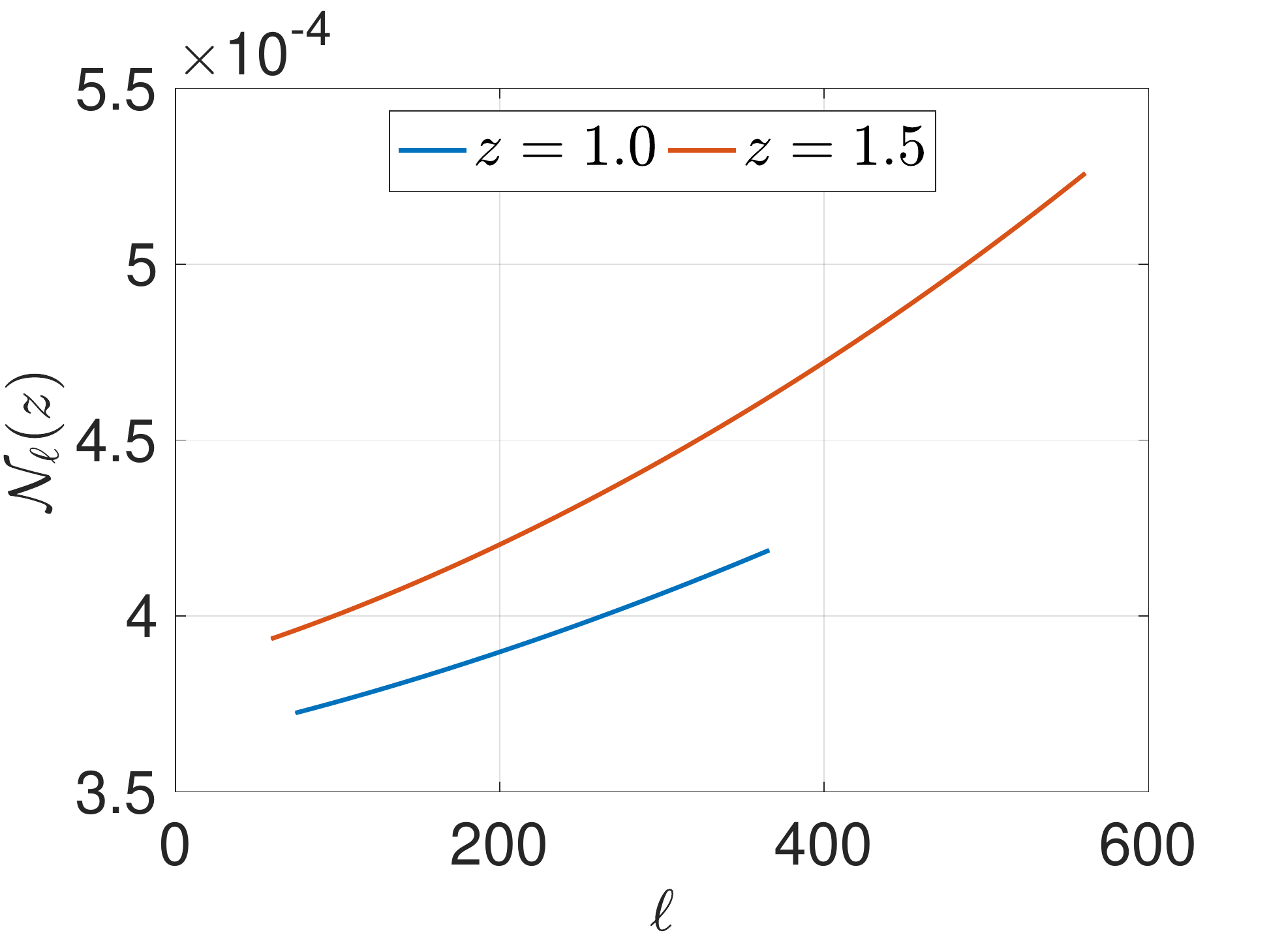}\\
\par\end{centering}
\caption{\label{fig:NoiExpts} 
SKA ({\em left}) and HIRAX ({\em right}) noise  at selected redshifts, {with $\Delta z=10^{-4}$}.    
}
\end{figure}

\noindent {The noise power spectra for SKA and HIRAX  surveys are shown at selected redshifts in Figure \ref{fig:NoiExpts}.}

\subsection{{Signal to noise of the bispectrum}}

The signal to noise ratio (SNR) for a fixed multipole configuration in a single redshift bin is 
\begin{equation}
\mathrm{SNR}_{\ell_1\ell_2\ell_3}(z)=\frac{\big|B_{\ell_1\ell_2\ell_3}(z)\big|}{\sigma_{B_{\ell_1\ell_2\ell_3}}(z)}\label{eq:SNR_Def}\,.
\end{equation}
 {The main contribution to the  variance, assuming {Gaussian initial conditions}, comes from the  Gaussian part of {the 6-point function,
 given by} \cite{DiDio:2018unb} 
\begin{equation}
   \sigma^2_{B_{\ell_1\ell_2\ell_3}}(z)={f_{\rm sky}^{-1}}\,
  {{\tilde  C_{\ell_1}(z)\,\tilde C_{\ell_2}(z)\,\tilde C_{\ell_3}(z)\, \big(1+2\delta_{\ell_1\ell_2}\delta_{\ell_2\ell_3}+\delta_{\ell_1\ell_2}+\delta_{\ell_2\ell_3}+\delta_{\ell_3\ell_1}\big)}\,,}\label{eq:variance_SNR}
\end{equation}
where 
\be
{\tilde  C_{\ell}(z)=  C_{\ell}(z)+{\cal N}_\ell(z)\,,}
\ee
and the noise is given by \eqref{genn}. We have generalised the expression in \cite{DiDio:2018unb} to include noise and to allow for $f_{\rm sky}\neq 1$. 
\begin{figure}[ht]
\begin{centering}
\includegraphics[scale=0.27]{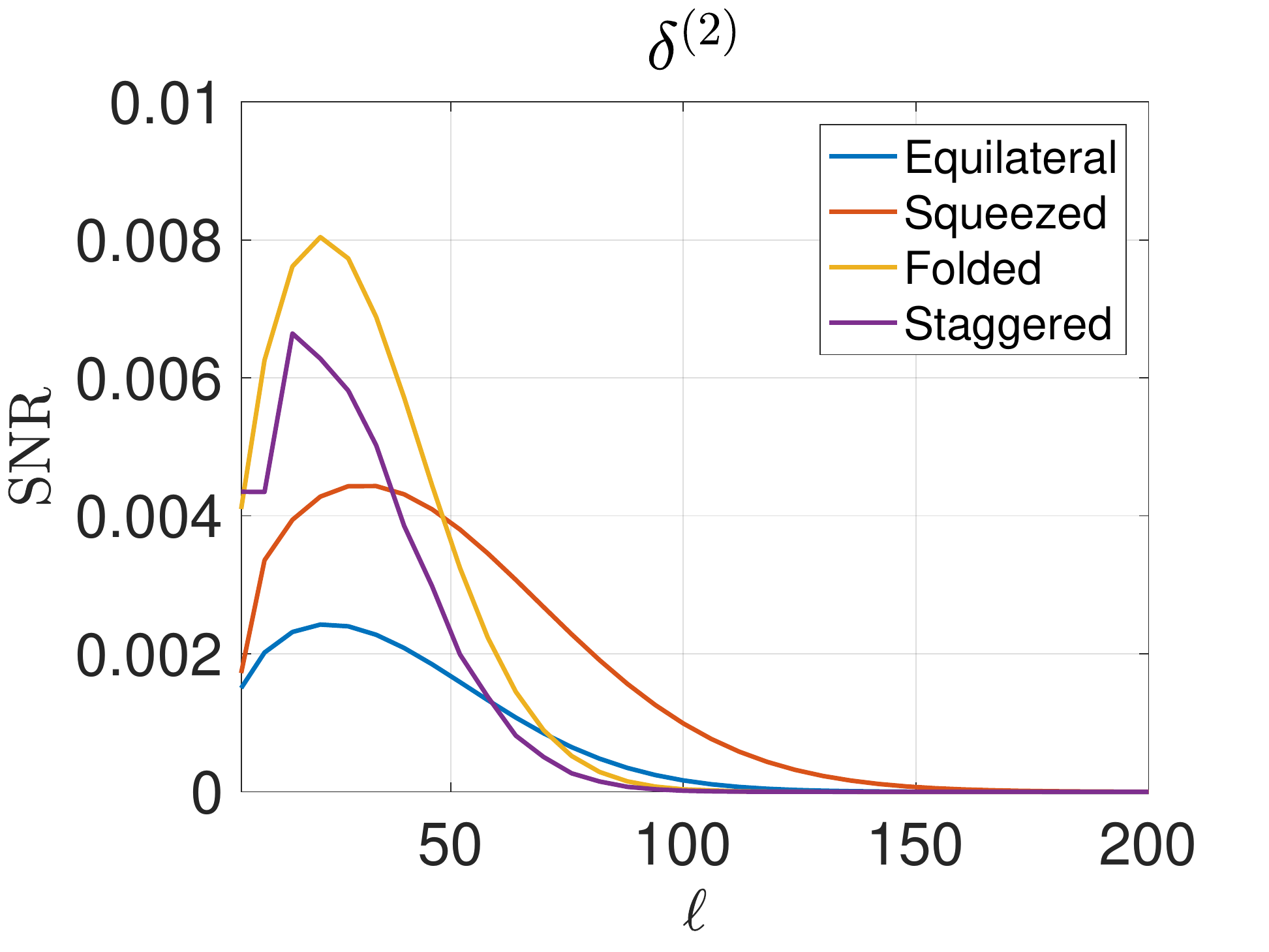}\includegraphics[scale=0.27]{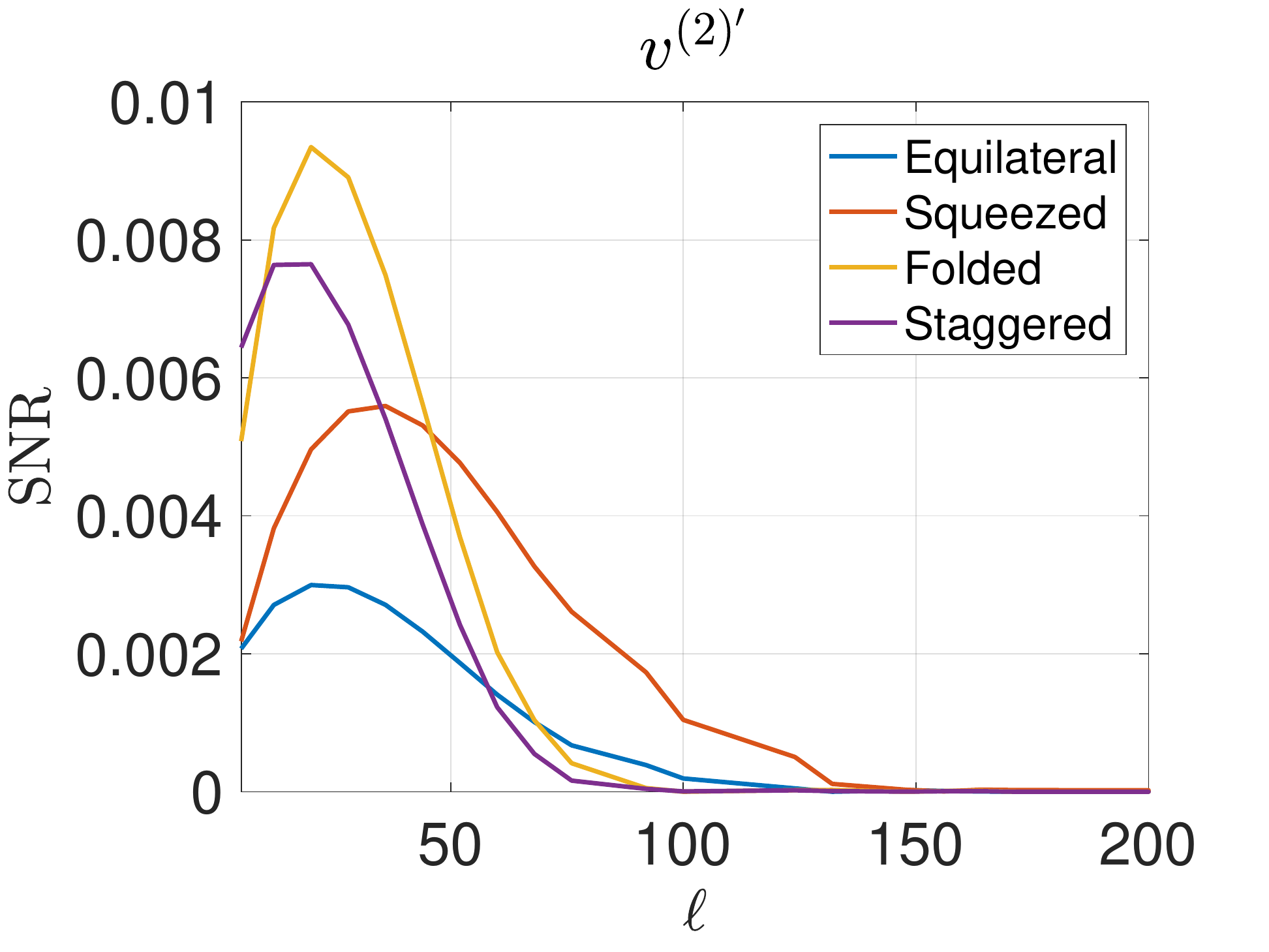}\includegraphics[scale=0.27]{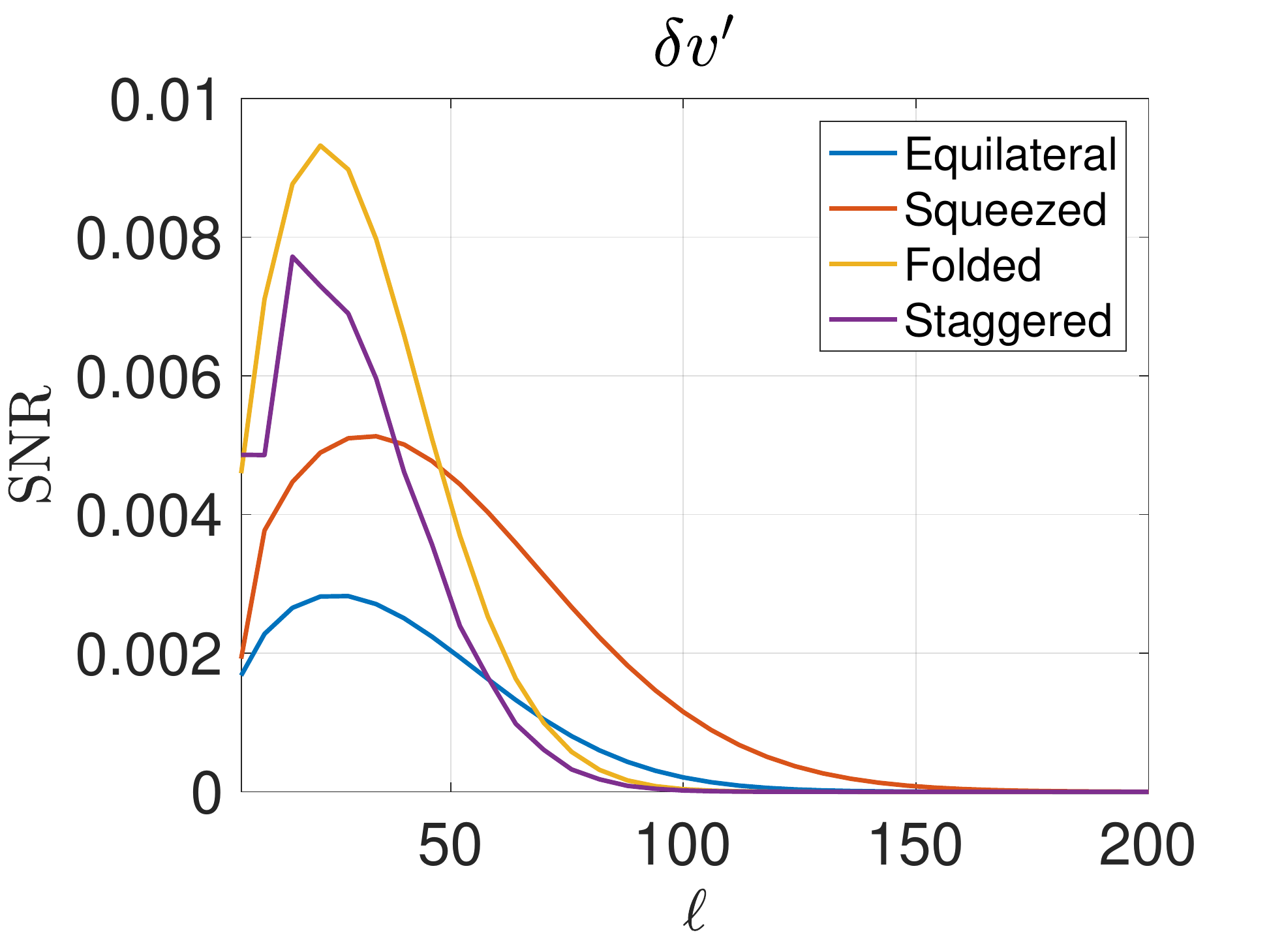} \\\includegraphics[scale=0.27]{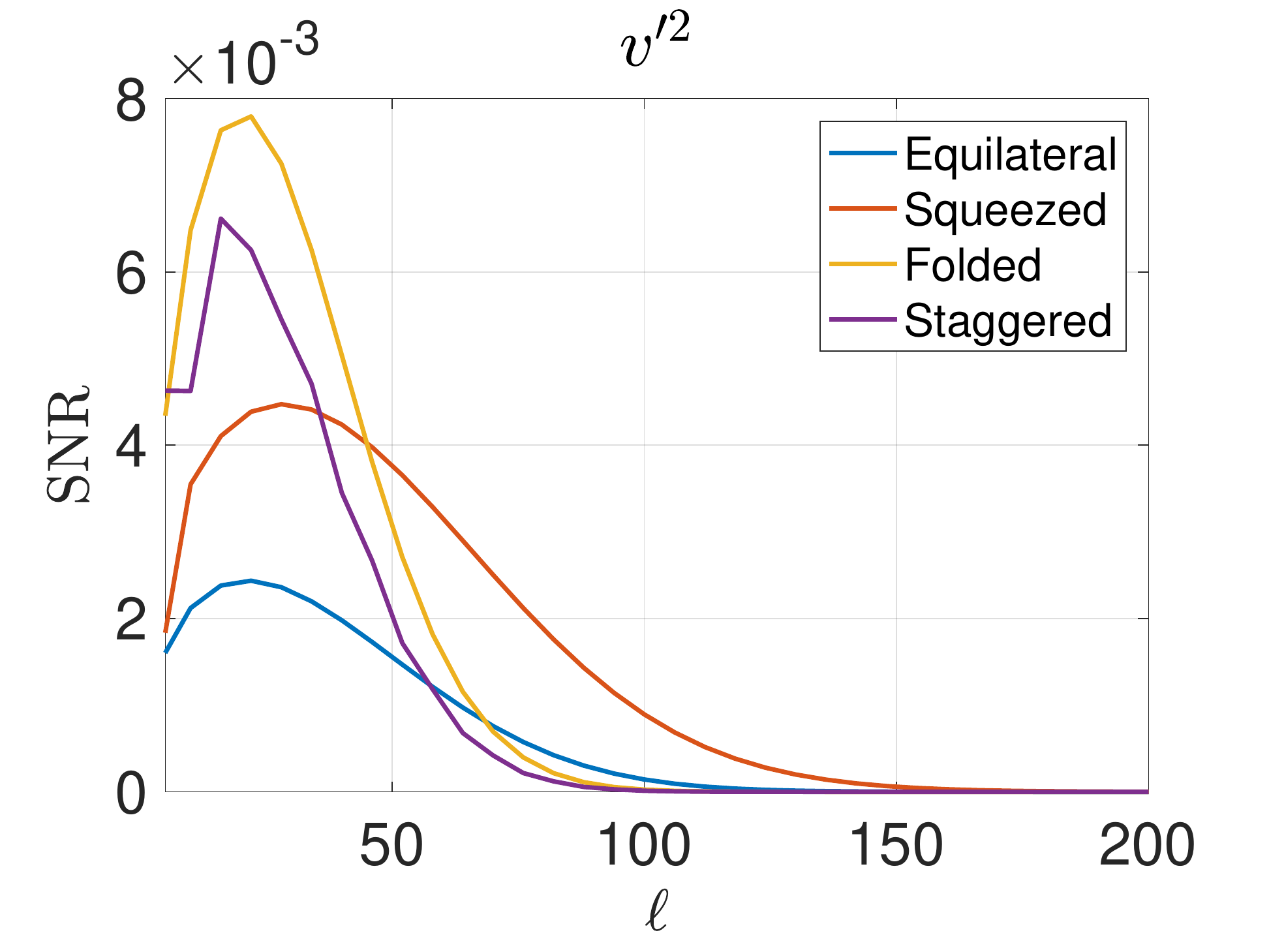}\includegraphics[scale=0.27]{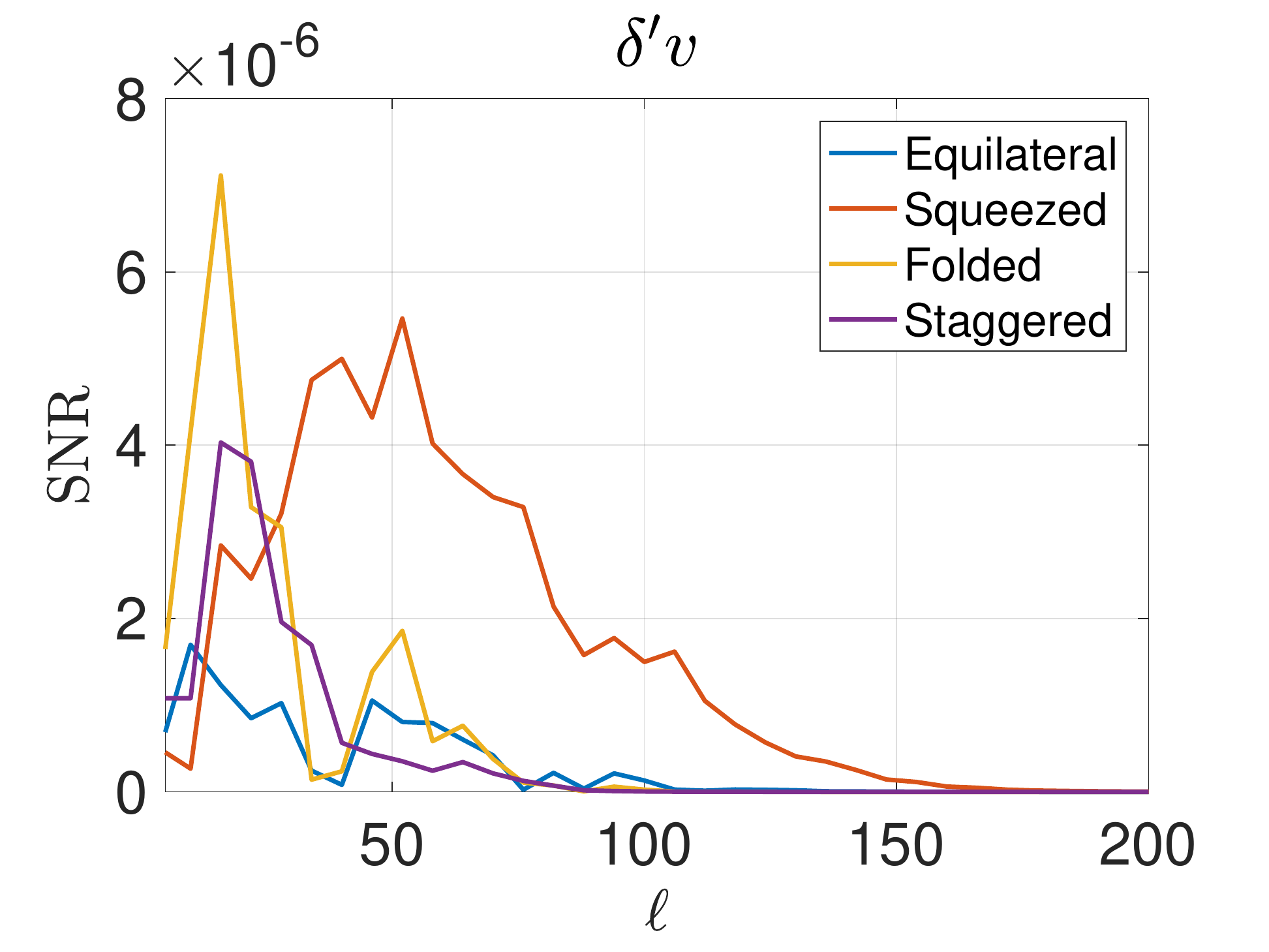}\includegraphics[scale=0.27]{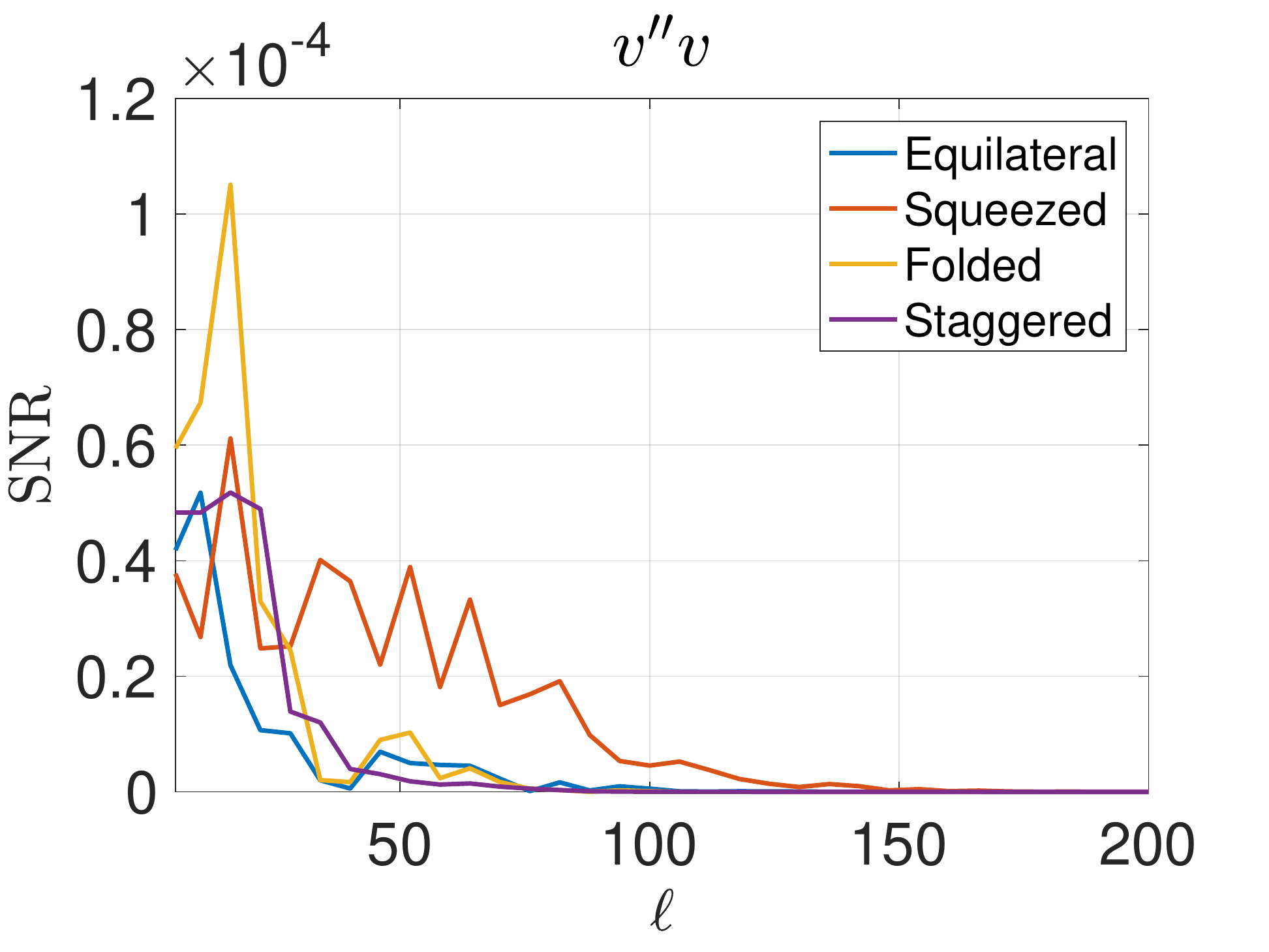}
\par\end{centering}
\caption{\label{fig:SNRBisp} {SNR per $\ell$ in one redshift bin for equilateral, squeezed ({$\ell_{\rm long}=5$}) flattened and staggered shapes for an SKA intensity map (SD mode) at {$z=1$}, with $\Delta z=10^{-4}$. The panels show the different contributions  in  \eqref{diff_bisp_cont}. {For the top middle panel we have used the filter of \cite{doi:10.1021/ac60214a047}, with order 4 and window size 15.}}}
\end{figure}

Figure \ref{fig:SNRBisp} shows the SNR per $\ell$ for the 3 shapes of Figure \ref{fig:RedBisp_Lims}, in the case of SKA in SD mode {(see assumption \ref{itm:approx_dirac} below for computation details)}. The different contributions of $B_{\ell_1\ell_2\ell_3}$ to the SNR, for equilateral, squeezed, flattened shapes and staggered configurations, are shown in the 6 panels.  As in the case of the theoretical signal, the contribution from the last two panels is subdominant. In all contributions, {SNR$_{\ell_1\ell_2\ell_3}$} becomes negligible for {$\ell\gtrsim 150$}, owing to the effect of the nonlinear cut-off: at $z=0.5$,  \eqref{lfgmin} gives $\ell_{\rm max}^{\rm \,nl}= 173$. {At higher redshifts, the growing contribution of the beam is what effectively cuts out the higher multipoles. This can be seen in Figure \ref{fig:LimScales}, where the beam resolution limit $\ell^{\rm \,res,SD}_{\rm max}$ replaces the nonlinear limit $\ell^{\rm \,nl}_{\rm max}$ as upper limit for $z\gtrsim 0.7$, and in Figure \ref{fig:NoiExpts} (left), where the noise power spectrum grows much more rapidly with $\ell$ for $z=1$ and 1.5 than for $z=0.5$.}

We need to compute the cumulative SNR in a redshift bin, summing over all multipoles and all shapes:
\begin{equation}
\mathrm{SNR}(z)^2={\sum_{{\ell_i}}\mathrm{SNR}_{\ell_1\ell_2\ell_3}(z)^2},\label{eq:CSNR_FORM}
\end{equation}
where the sum is over all triangular configurations obtained after imposing the  Wigner 3j conditions:
(a) $\ell_1+\ell_2+\ell_3$ is even, (b) $|\ell_1-\ell_2|\le\ell_3\le\ell_1+\ell_2$. We also use $\ell_1\le\ell_2\le\ell_3$, exploiting invariance of the bispectrum under multipole permutations.

In order to demonstrate the detectability of the 21cm bispectrum, we only require a rough estimate of the cumulative SNR. With this in mind, we make three simplifying assumptions to speed up the computations:
\begin{enumerate}[label=(A\arabic*), ref=(A\arabic*), font=\bfseries] 
\item \label{itm:approx_rsd}
{We neglect the second-order RSD contribution to the signal, including only the linear RSD contribution. 
\begin{figure}[!ht]
\begin{centering}
\includegraphics[scale=0.5]{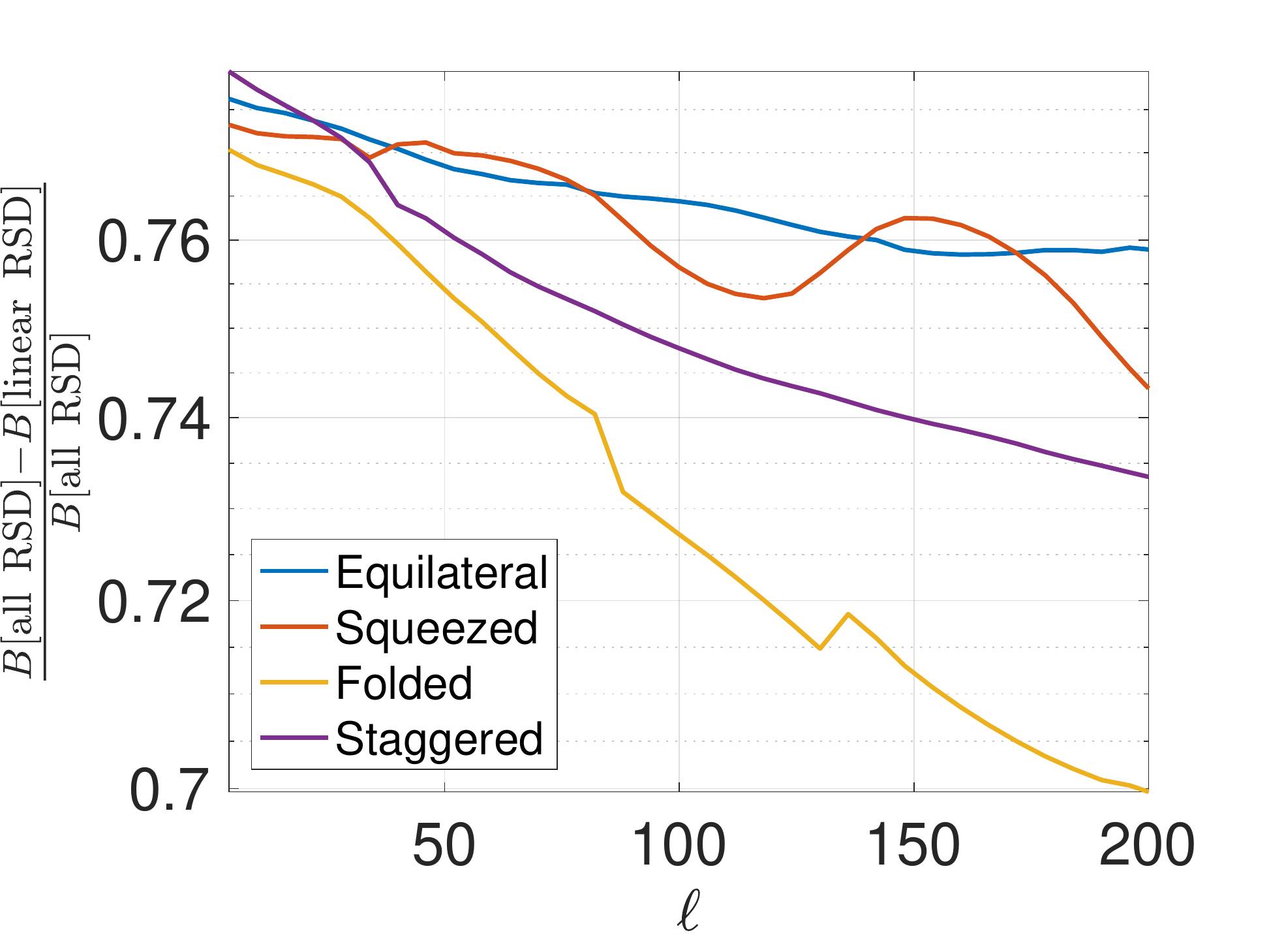}
\par\end{centering}
\caption{\label{linrsd} As in Figure  \ref{fig:tot_bisp}, but showing here the fractional contribution  \eqref{dbb2} of linear RSD to the full bispectrum.}
\end{figure}

{In other words, we use 
\\ \hspace*{3cm}{$ B[\mbox{linear RSD}]= \big \langle \Delta^{(1)}_1\Delta^{(1)}_2{\big({\Delta}_3^{(2)}-R_3^{(2)}\big)}\big\rangle + \mathrm{2 \ perms}$,}\\ {where $R^{(2)}$ denotes all the RSD terms in \eqref{eq:delta_sec_ord},} instead of\\ \hspace*{3.4cm}{ $ B[\mbox{all RSD}]= \big \langle \Delta^{(1)}_1\Delta^{(1)}_2{\Delta}_3^{(2)}\big\rangle + \mathrm{2 \ perms}$.}\\
Figure \ref{linrsd} shows the fractional contribution
\be\label{dbb2}
B[\mbox{all RSD}]-B[\mbox{linear RSD}] \over B[\mbox{all RSD}]\,
\ee
of the the linear RSD to the full bispectrum for the three shapes, similar to Figure \ref{fig:tot_bisp}. From Figure \ref{linrsd}, we see that the linear-only RSD signal is $\sim 25\%$ of the full signal. This means that we under-estimate the signal, and therefore the SNR, by a factor of $\sim 4$, when we include only the linear RSD and exclude the nonlinear RSD contribution.}
{We therefore multiply the SNR calculated from $B[\mbox{linear RSD}]$ by a factor of 4, as a rough estimate for the full SNR from $B[\mbox{all RSD}]$.} 
It is interesting to note that while the folded triangles give the largest total signal, the contribution from the non-linear RSD is somewhat smaller for this configuration than for the equilateral, squeezed and staggered ones. It is between 76\% and 70\% while for the three 
other configurations it never drops below 73.5\%. This is due to the fact that the linear RSD contributes the most to the total signal in the folded configuration (see Fig.~\ref{fig:All_Bisp}).
\item \label{itm:approx_snr}
The number of possible triangular configurations {with non-vanishing Wigner $3j$-symbols} rises rapidly as  $\ell_{\mathrm{max}}$ increases. We use the simple approximation proposed in \cite{Montanari:2020}:
\be\label{snr*}
\mathrm{SNR}_*(z)^2= n_\Delta\, \overline{{\rm SNR}_{\ell_i}(z)^2}\,,
\ee
where $n_\Delta$ is the total number of triangles and the arithmetic mean $\overline{{\rm SNR}^2_{\ell_i}}$ is estimated by computing the SNR for a random selection of triangles.

    \item \label{itm:approx_dirac}
We use a Dirac delta window {(see section \ref{sec:win})} to compute the signal, thus avoiding the numerical complexities of applying a window function in angular redshift space. 

For the smallest possible redshift bin-width for next-generation 21cm intensity maps, we take   $\Delta z = 10^{-4}$, and we use this in the thermal noise \eqref{genn}. Effectively, this assumes that the signal with $\Delta z=0$  is approximately the same as the signal with $\Delta z = 10^{-4}$. }
\end{enumerate}

\noindent The results based on \ref{itm:approx_rsd}--\ref{itm:approx_dirac}
 are shown  in Tables \ref{tab:SKACSNR} and \ref{tab:HiraxCSNR}. {As expected, smaller redshifts lead to larger SNR due to larger non-Gaussianities induced by the non-linear gravitational evolution. While the largest SNRs reported here are already promising for single-bin detection, the ${\rm SNR} \lesssim 1$ cases are also relevant in view of tomographic analysis that can potentially benefit from the joint signal of hundreds of bins and their cross-correlations.}

\begin{center}
\begin{table}[!ht]
\begin{centering}
\begin{tabular}{c c c c c c}
\hline 
\hline 
\noalign{\vskip 0.1cm}
Redshift & $\ell_{\mathrm{min}}$ & $\ell_{\mathrm{max}}$ & $n_\Delta$ & SNR & SNR \\
&&&&{(linear RSD)}  & {(all RSD, estimated)} \\
 \noalign{\vskip 0.1cm}
\hline
\hline
\noalign{\vskip 0.1cm}
\noalign{\vskip 0.1cm}
0.5 & 5 & 173 & 225598 & 5.04 & {$\sim$20\,\,}\\
1.0 & 5 & 224 & 485380 & 0.69 & {$\sim$3~\,\,}  \\
1.5 & 5 & 179 & 249576 & {0.09} & {$\sim$0.4} \\
\hline 
\hline 
\noalign{\vskip 0.1cm}
\end{tabular}
\par\end{centering}
\caption{\label{tab:SKACSNR} Cumulative SNR in one redshift bin for an SKA survey: {using \eqref{snr*} for the linear RSD case, and roughly estimating the full linear + nonlinear RSD contribution by applying a factor 4, as described in \ref{itm:approx_rsd} above (see Figure \ref{linrsd}).} }
\end{table}
\begin{table}[h!]
\begin{centering}
\begin{tabular}{c c c c c c}
\hline 
\hline 
\noalign{\vskip 0.1cm}
 Redshift & $\ell_{\mathrm{min}}$ & $\ell_{\mathrm{max}}$ & $n_\Delta$ & SNR & SNR\\
&&&&{(linear RSD)}  & {(all RSD, estimated)} \\
 \noalign{\vskip 0.1cm}
\hline
\hline
\noalign{\vskip 0.1cm}
1.0 & 74 & 366 & 1680096 & 2.78 & {$\sim$11}\\
\noalign{\vskip 0.1cm}
1.5 & 59 & 561 & 7021562 & {1.08} & {$\sim$4~}\\
\hline 
\hline 
\noalign{\vskip 0.1cm}
\end{tabular}
\par\end{centering}
\caption{\label{tab:HiraxCSNR} 
{As in Table \ref{tab:SKACSNR},} for  a HIRAX survey.
}
\end{table}
\par\end{center}

\subsection{Consistency checks}
\label{sec:win}

In this section, we assess the {validity} of approximations \ref{itm:approx_snr} and \ref{itm:approx_dirac} above}. We consider the SKA SD case at $z = 0.5$. 

\begin{figure}[!ht]
    \centering
    \includegraphics[width=0.49\textwidth]{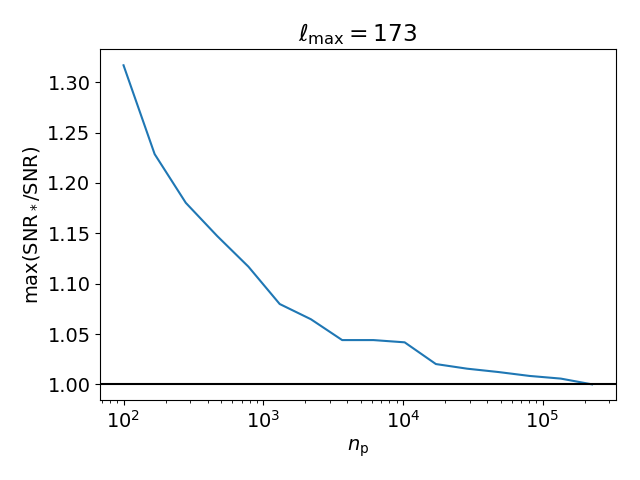}
    \includegraphics[width=0.49\textwidth]{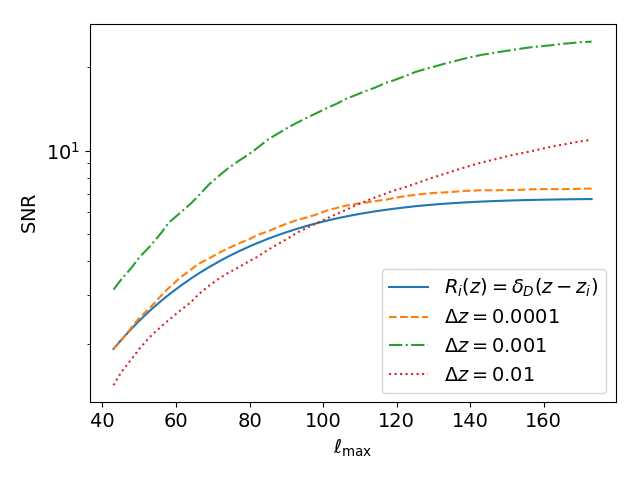}
    \caption{{Here we assume} an SKA survey at $z =0.5$.
    \emph{Left:} Estimated error induced by the approximate cumulative ${{\rm SNR}_*}$ in \eqref{snr*}, obtained by considering only a partial subset of $n_{\rm p}$ multipole triangles. For each $n_{\rm p}$, we consider 1000 random selections of $n_{\rm p}$ multipole triangles and  plot the largest deviation with respect to the non-approximate cumulative SNR result \eqref{eq:CSNR_FORM}. \emph{Right:} {Approximate} cumulative SNR as a function of the maximum multipole $\ell_{\rm max}$ for single redshift bins of different widths.}
    \label{fig:win}
\end{figure}

For \ref{itm:approx_snr}, we  estimate the error in SNR induced by approximating the arithmetic mean $\overline{{\rm SNR}_{\ell_i}^2}$ in \eqref{snr*}, using only a partial subset of $n_{\rm p}$ multipole triangles, instead of all $n_{\Delta}$ possible triangles. The left panel of Figure \ref{fig:win} shows the deviation with respect to the full result. The latter is obtained {by} summing over all $n_{\Delta}$ triangles that we compute up to $\ell_{\rm max}=173$. To mitigate the risk of under-estimating the error due to the choice of a particular random draw of $n_{\rm p}$ triangles, we consider 1000 random draws for each $n_{\rm p}$ and compute the maximum cumulative SNR deviation over all draws. From Figure \ref{fig:win}, we expect $\lesssim 10\%$ errors when considering $\gtrsim 10^3$ triangles.

Next, we check assumption \ref{itm:approx_dirac}, i.e., whether neglecting the numerically expensive integration over redshift bins recovers the cumulative SNR for the small $\Delta z= 10^{-4}$ used in our forecasts. The right panel of Figure \ref{fig:win} shows the {approximate} cumulative SNR as a function of $\ell_{\rm max}$ for different redshift bin widths.
The case labeled $R_i(z) = \delta_D(z-z_i)$ corresponds to approximating radial selection functions (see below) as Dirac deltas -- i.e., effectively neglecting integration over redshifts, as in our forecasts. For this case, we include a finite $\Delta z = 10^{-4}$ in the noise term via \eqref{eq:Dnu}. The other cases consistently integrate over tophat radial selection functions ${R_i}(z)$ of width $\Delta z > 0$:
\bea
  \label{eq:14}
  B_{\ell_1\ell_2\ell_3}^{ijk} &=& \int{\rm d}z_1\, R_i(z_1) \int{\rm d}z_2\, R_j(z_2) \int{\rm d}z_3 \,R_k(z_3)\,
   B_{\ell_1\ell_2\ell_3} (z_1, z_2, z_3)\,,
\\
  \label{eq:17}
  \sigma^2_{B_{\ell_1\ell_2\ell_3}^{ijk}} &=& \int{\rm d}z_1\, R_i(z_1) \int{\rm d}z_2\, R_j(z_2) \int{\rm d}z_3\, R_k(z_3)\, \sigma^2_{B_{\ell_1\ell_2\ell_3}}(z_1, z_2, z_3) \,.
\eea
The variance is given by
\begin{eqnarray} \label{eq:B_covarince_sumleven}
  {f_{\rm sky}}\,\sigma^2_{B_{\ell_1\ell_2\ell_3}}(z_1, z_2, z_3) &=&
     \tilde C_{\ell_1}^{11}  \tilde C_{\ell_2}^{22}
     \tilde C_{\ell_3}^{33}
     + \left[
     \tilde C_{\ell_1}^{12} \tilde  C_{\ell_2}^{23}
      \tilde C_{\ell_3}^{31}
     +
     \tilde C_{\ell_1}^{13}  \tilde C_{\ell_2}^{21}
      \tilde C_{\ell_3}^{32}  \right]
     \delta_{\ell_1\ell_2} \delta_{\ell_2\ell_3}
     \nonumber \\
  &&{}  +
     \tilde C_{\ell_1}^{11}  \tilde C_{\ell_2}^{23}
      \tilde C_{\ell_3}^{32}
     \delta_{\ell_2\ell_3}
     +
     \tilde C_{\ell_1}^{12} \tilde  C_{\ell_2}^{21}
      \tilde C_{\ell_3}^{33}
     \delta_{\ell_1\ell_2} +
     \tilde C_{\ell_1}^{13} \tilde  C_{\ell_2}^{22}
      \tilde C_{\ell_3}^{31}
     \delta_{\ell_1\ell_3} \,,~~
\end{eqnarray}
where $\tilde C_{\ell}^{ij} = C_{\ell}(z_i, z_j) + {\cal N}_\ell(z_i) \delta_{ij}$. {We estimate SNR$_*$ by summing over $n_{\rm p} \sim 2000$ triangles for each $\Delta z > 0$ case. The left panel of Figure \ref{fig:win} then suggests $\lesssim 10\%$ methodological errors.} 

Our forecast approximation converges  well to the case $\Delta z = 10^{-4}$, which validates our analysis. The redshift bin of width $\Delta z=10^{-3}$ leads to a larger signal for the single redshift bin under consideration, given the smaller impact of the noise. However, smoothing the signal over the even larger $\Delta z=10^{-2}$ degrades the SNR as it reduces the signal more strongly than the noise. 
{Note that these individual-bin SNR do not give an accurate picture of the full SNR:}
 smaller $\Delta z$ values allow for a finer tomographic reconstruction, {i.e., a larger number of redshift bins -- which increases the SNR. We leave the detailed study of an optimal binning strategy} as a future development of our work. 

\section{Conclusions}
\label{sec:conclusions}

In this work, we computed the tree-level bispectrum of 21cm intensity mapping induced at second order in  perturbations. We  worked in directly observable angle-redshift space, which includes all wide-angle, RSD  effects. As we discussed  equal redshift bins, $z_1=z_2=z_3$, we have neglected lensing which is subdominant in these configurations as shown in~\cite{DiDio:2015bua}. The bispectrum is dominated by the second-order RSD contributions. We expect this finding to be valid also for spectroscopic number count bispectra {that also allows bispectrum estimation in relatively small redshift bins.} 

{We  computed the SNR for the two near-future surveys SKA-MID (single-dish mode) and HIRAX (interferometric mode). We found that  ${\rm SNR} \gtrsim 10$ can be reached in a single bin of width $\Delta z=10^{-4}$ for SKA at redshift $z=0.5$ and for HIRAX at $z=1$.  
At other redshifts studied in detail, the single-redshift SNR is less than 5 and several bins need to be combined in order to reach an SNR of 10. Another possibility is to increase the bin width. For example, increasing $\Delta z=10^{-4}$  to $\Delta z=10^{-3}$ at {$z=0.5$} leads to an increase of the SNR by a factor of  {$\sim$2}. This is, however less than the number of independent bins of width $10^{-4}$ that we could place inside a $\Delta z=10^{-3}$ bin -- and the SNR would be further increased by cross correlations. The optimal binning strategy, in both redshift and multipole space, in order to achieve the best SNR, will depend on the detailed observations and is left for a future project. Here we have shown that the detection of the bispectrum with next-generation radio telescopes is feasible.}
{It will also be  interesting to investigate whether the SNR of the bispectrum from significantly different redshift bins, where the lensing term is dominant~\cite{DiDio:2015bua}, is sufficient to allow its detection. We leave this for a future investigation.}

{Interesting theoretical questions raised include:  How does this `guaranteed' bispectrum, which is a consequence of the nonlinearity of gravity on Gaussian initial perturbations, compare with a possible primordial bispectrum from inflation \cite{DiDio:2016gpd}?  How does its shape compare with the simple $f_{\rm NL}$ non-Gaussianity expected in many inflationary models?}

Naively, we expect an inflationary $f_{\rm NL}$ to dominate in the squeezed configuration, which is not the case for the bispectrum from nonlinearities investigated here. Therefore, the distinction between these contributions might not be too difficult to detect if the SNR of the experiment and the $f_{\rm NL}$ of the model are sufficiently large.

{Another avenue for future investigations is the question of how the additional information in the bispectrum can improve constraints on cosmological model parameters.}

\acknowledgments
{We thank Enea Di Dio for pointing out an error in our original analysis of the lensing contribution.}
RK is  extremely grateful to Sandeep Sirothia for assistance with software technicalities.
{FM thanks Stefano Camera for useful discussions.}
RD and MJ acknowledge support from the Swiss National Science Foundation.
RK and RM are supported by the South African Radio Astronomy Observatory and the National Research Foundation (Grant No. 75415). RM is also supported by the UK Science \& Technology Facilities Council (Grant ST/N000668/1). FM is supported by the Research Project FPA2015-68048-C3-3-P [MINECO-FEDER]
and the Centro de Excelencia Severo Ochoa Program SEV-2016-0597.
This work  made use of the South African Centre for High Performance Computing, under the project {\em Cosmology with Radio Telescopes,} ASTRO-0945, and of the Kerbero cluster at IFT-UAM/CSIC (Madrid, Spain).

\clearpage
\bibliographystyle{JHEP}
\bibliography{bibliography}

\providecommand{\href}[2]{#2}\begingroup\raggedright\begin{thebibliography}{10}

\bibitem{Fonseca:2016qqw}
J.~Fonseca, M.~Silva, M.~G. Santos and A.~Cooray, \emph{{Cosmology with
  intensity mapping techniques using atomic and molecular lines}},
  \href{https://doi.org/10.1093/mnras/stw2470}{\emph{Mon. Not. Roy. Astron.
  Soc.} {\bfseries 464} (2017) 1948}
  [\href{https://arxiv.org/abs/1607.05288}{{\ttfamily 1607.05288}}].

\bibitem{Kovetz:2017agg}
E.~D. Kovetz et~al., \emph{{Line-Intensity Mapping: 2017 Status Report}},
  \href{https://arxiv.org/abs/1709.09066}{{\ttfamily 1709.09066}}.

\bibitem{Hall:2012wd}
A.~Hall, C.~Bonvin and A.~Challinor, \emph{{Testing General Relativity with
  21-cm intensity mapping}},
  \href{https://doi.org/10.1103/PhysRevD.87.064026}{\emph{Phys. Rev. D}
  {\bfseries 87} (2013) 064026}
  [\href{https://arxiv.org/abs/1212.0728}{{\ttfamily 1212.0728}}].

\bibitem{Bull:2014rha}
P.~Bull, P.~G. Ferreira, P.~Patel and M.~G. Santos, \emph{{Late-time cosmology
  with 21cm intensity mapping experiments}},
  \href{https://doi.org/10.1088/0004-637X/803/1/21}{\emph{Astrophys. J.}
  {\bfseries 803} (2015) 21} [\href{https://arxiv.org/abs/1405.1452}{{\ttfamily
  1405.1452}}].

\bibitem{Bacon:2018dui}
{\scshape SKA} collaboration, \emph{{Cosmology with Phase 1 of the Square
  Kilometre Array: Red Book 2018: Technical specifications and performance
  forecasts}}, \href{https://doi.org/10.1017/pasa.2019.51}{\emph{Publ. Astron.
  Soc. Austral.} {\bfseries 37} (2020) e007}
  [\href{https://arxiv.org/abs/1811.02743}{{\ttfamily 1811.02743}}].

\bibitem{Fonseca:2019qek}
J.~Fonseca, J.-A. Viljoen and R.~Maartens, \emph{{Constraints on the growth
  rate using the observed galaxy power spectrum}},
  \href{https://doi.org/10.1088/1475-7516/2019/12/028}{\emph{JCAP} {\bfseries
  12} (2019) 028} [\href{https://arxiv.org/abs/1907.02975}{{\ttfamily
  1907.02975}}].

\bibitem{Alonso:2015uua}
D.~Alonso, P.~Bull, P.~G. Ferreira, R.~Maartens and M.~Santos, \emph{{Ultra
  large-scale cosmology in next-generation experiments with single tracers}},
  \href{https://doi.org/10.1088/0004-637X/814/2/145}{\emph{Astrophys. J.}
  {\bfseries 814} (2015) 145}
  [\href{https://arxiv.org/abs/1505.07596}{{\ttfamily 1505.07596}}].

\bibitem{Fonseca:2015laa}
J.~Fonseca, S.~Camera, M.~Santos and R.~Maartens, \emph{{Hunting down
  horizon-scale effects with multi-wavelength surveys}},
  \href{https://doi.org/10.1088/2041-8205/812/2/L22}{\emph{Astrophys. J. Lett.}
  {\bfseries 812} (2015) L22}
  [\href{https://arxiv.org/abs/1507.04605}{{\ttfamily 1507.04605}}].

\bibitem{Umeh:2015gza}
O.~Umeh, R.~Maartens and M.~Santos, \emph{{Nonlinear modulation of the HI power
  spectrum on ultra-large scales. I}},
  \href{https://doi.org/10.1088/1475-7516/2016/03/061}{\emph{JCAP} {\bfseries
  03} (2016) 061} [\href{https://arxiv.org/abs/1509.03786}{{\ttfamily
  1509.03786}}].

\bibitem{Jalivand:2018vfz}
M.~Jalilvand, E.~Majerotto, R.~Durrer and M.~Kunz, \emph{{Intensity mapping of
  the 21 cm emission: lensing}},
  \href{https://doi.org/10.1088/1475-7516/2019/01/020}{\emph{JCAP} {\bfseries
  01} (2019) 020} [\href{https://arxiv.org/abs/1807.01351}{{\ttfamily
  1807.01351}}].

\bibitem{DiDio:2015bua}
E.~Di~Dio, R.~Durrer, G.~Marozzi and F.~Montanari, \emph{{The bispectrum of
  relativistic galaxy number counts}},
  \href{https://doi.org/10.1088/1475-7516/2016/01/016}{\emph{JCAP} {\bfseries
  01} (2016) 016} [\href{https://arxiv.org/abs/1510.04202}{{\ttfamily
  1510.04202}}].

\bibitem{DiDio:2018unb}
E.~Di~Dio, R.~Durrer, R.~Maartens, F.~Montanari and O.~Umeh, \emph{{The
  Full-Sky Angular Bispectrum in Redshift Space}},
  \href{https://doi.org/10.1088/1475-7516/2019/04/053}{\emph{JCAP} {\bfseries
  04} (2019) 053} [\href{https://arxiv.org/abs/1812.09297}{{\ttfamily
  1812.09297}}].

\bibitem{Umeh:2016nuh}
O.~Umeh, S.~Jolicoeur, R.~Maartens and C.~Clarkson, \emph{{A general
  relativistic signature in the galaxy bispectrum: the local effects of
  observing on the lightcone}},
  \href{https://doi.org/10.1088/1475-7516/2017/03/034}{\emph{JCAP} {\bfseries
  03} (2017) 034} [\href{https://arxiv.org/abs/1610.03351}{{\ttfamily
  1610.03351}}].

\bibitem{Yoo:2010ni}
J.~Yoo, \emph{{General Relativistic Description of the Observed Galaxy Power
  Spectrum: Do We Understand What We Measure?}},
  \href{https://doi.org/10.1103/PhysRevD.82.083508}{\emph{Phys. Rev. D}
  {\bfseries 82} (2010) 083508}
  [\href{https://arxiv.org/abs/1009.3021}{{\ttfamily 1009.3021}}].

\bibitem{Challinor:2011bk}
A.~Challinor and A.~Lewis, \emph{{The linear power spectrum of observed source
  number counts}},
  \href{https://doi.org/10.1103/PhysRevD.84.043516}{\emph{Phys. Rev. D}
  {\bfseries 84} (2011) 043516}
  [\href{https://arxiv.org/abs/1105.5292}{{\ttfamily 1105.5292}}].

\bibitem{Bonvin:2011bg}
C.~Bonvin and R.~Durrer, \emph{{What galaxy surveys really measure}},
  \href{https://doi.org/10.1103/PhysRevD.84.063505}{\emph{Phys. Rev. D}
  {\bfseries 84} (2011) 063505}
  [\href{https://arxiv.org/abs/1105.5280}{{\ttfamily 1105.5280}}].

\bibitem{Bertacca:2014dra}
D.~Bertacca, R.~Maartens and C.~Clarkson, \emph{{Observed galaxy number counts
  on the lightcone up to second order: I. Main result}},
  \href{https://doi.org/10.1088/1475-7516/2014/09/037}{\emph{JCAP} {\bfseries
  09} (2014) 037} [\href{https://arxiv.org/abs/1405.4403}{{\ttfamily
  1405.4403}}].

\bibitem{Yoo:2014sfa}
J.~Yoo and M.~Zaldarriaga, \emph{{Beyond the Linear-Order Relativistic Effect
  in Galaxy Clustering: Second-Order Gauge-Invariant Formalism}},
  \href{https://doi.org/10.1103/PhysRevD.90.023513}{\emph{Phys. Rev. D}
  {\bfseries 90} (2014) 023513}
  [\href{https://arxiv.org/abs/1406.4140}{{\ttfamily 1406.4140}}].

\bibitem{DiDio:2014lka}
E.~Di~Dio, R.~Durrer, G.~Marozzi and F.~Montanari, \emph{{Galaxy number counts
  to second order and their bispectrum}},
  \href{https://doi.org/10.1088/1475-7516/2014/12/017}{\emph{JCAP} {\bfseries
  12} (2014) 017} [\href{https://arxiv.org/abs/1407.0376}{{\ttfamily
  1407.0376}}].

\bibitem{Kehagias:2015tda}
A.~Kehagias, A.~Moradinezhad~Dizgah, J.~Noreña, H.~Perrier and A.~Riotto,
  \emph{{A Consistency Relation for the Observed Galaxy Bispectrum and the
  Local non-Gaussianity from Relativistic Corrections}},
  \href{https://doi.org/10.1088/1475-7516/2015/08/018}{\emph{JCAP} {\bfseries
  08} (2015) 018} [\href{https://arxiv.org/abs/1503.04467}{{\ttfamily
  1503.04467}}].

\bibitem{Jolicoeur:2017nyt}
S.~Jolicoeur, O.~Umeh, R.~Maartens and C.~Clarkson, \emph{{Imprints of local
  lightcone projection effects on the galaxy bispectrum. II}},
  \href{https://doi.org/10.1088/1475-7516/2017/09/040}{\emph{JCAP} {\bfseries
  1709} (2017) 040} [\href{https://arxiv.org/abs/1703.09630}{{\ttfamily
  1703.09630}}].

\bibitem{Jolicoeur:2017eyi}
S.~Jolicoeur, O.~Umeh, R.~Maartens and C.~Clarkson, \emph{{Imprints of local
  lightcone projection effects on the galaxy bispectrum. Part III. Relativistic
  corrections from nonlinear dynamical evolution on large-scales}},
  \href{https://doi.org/10.1088/1475-7516/2018/03/036}{\emph{JCAP} {\bfseries
  1803} (2018) 036} [\href{https://arxiv.org/abs/1711.01812}{{\ttfamily
  1711.01812}}].

\bibitem{Jolicoeur:2018blf}
S.~Jolicoeur, A.~Allahyari, C.~Clarkson, J.~Larena, O.~Umeh and R.~Maartens,
  \emph{{Imprints of local lightcone projection effects on the galaxy
  bispectrum IV: Second-order vector and tensor contributions}},
  \href{https://doi.org/10.1088/1475-7516/2019/03/004}{\emph{JCAP} {\bfseries
  03} (2019) 004} [\href{https://arxiv.org/abs/1811.05458}{{\ttfamily
  1811.05458}}].

\bibitem{Clarkson:2018dwn}
C.~Clarkson, E.~M. de~Weerd, S.~Jolicoeur, R.~Maartens and O.~Umeh, \emph{{The
  dipole of the galaxy bispectrum}},
  \href{https://doi.org/10.1093/mnrasl/slz066}{\emph{Mon. Not. Roy. Astron.
  Soc.} {\bfseries 486} (2019) L101}
  [\href{https://arxiv.org/abs/1812.09512}{{\ttfamily 1812.09512}}].

\bibitem{Maartens:2019yhx}
R.~Maartens, S.~Jolicoeur, O.~Umeh, E.~M. De~Weerd, C.~Clarkson and S.~Camera,
  \emph{{Detecting the relativistic galaxy bispectrum}},
  \href{https://doi.org/10.1088/1475-7516/2020/03/065}{\emph{JCAP} {\bfseries
  03} (2020) 065} [\href{https://arxiv.org/abs/1911.02398}{{\ttfamily
  1911.02398}}].

\bibitem{deWeerd:2019cae}
E.~M. de~Weerd, C.~Clarkson, S.~Jolicoeur, R.~Maartens and O.~Umeh,
  \emph{{Multipoles of the relativistic galaxy bispectrum}},
  \href{https://doi.org/10.1088/1475-7516/2020/05/018}{\emph{JCAP} {\bfseries
  05} (2020) 018} [\href{https://arxiv.org/abs/1912.11016}{{\ttfamily
  1912.11016}}].

\bibitem{Desjacques:2016bnm}
V.~Desjacques, D.~Jeong and F.~Schmidt, \emph{{Large-Scale Galaxy Bias}},
  \href{https://doi.org/10.1016/j.physrep.2017.12.002}{\emph{Phys. Rept.}
  {\bfseries 733} (2018) 1} [\href{https://arxiv.org/abs/1611.09787}{{\ttfamily
  1611.09787}}].

\bibitem{Nielsen:2016ldx}
J.~T. Nielsen and R.~Durrer, \emph{{Higher order relativistic galaxy number
  counts: dominating terms}},
  \href{https://doi.org/10.1088/1475-7516/2017/03/010}{\emph{JCAP} {\bfseries
  03} (2017) 010} [\href{https://arxiv.org/abs/1606.02113}{{\ttfamily
  1606.02113}}].

\bibitem{Komatsu:2001rj}
E.~Komatsu and D.~N. Spergel, \emph{{Acoustic signatures in the primary
  microwave background bispectrum}},
  \href{https://doi.org/10.1103/PhysRevD.63.063002}{\emph{Phys. Rev. D}
  {\bfseries 63} (2001) 063002}
  [\href{https://arxiv.org/abs/astro-ph/0005036}{{\ttfamily
  astro-ph/0005036}}].

\bibitem{Bernardeau:2001qr}
F.~Bernardeau, S.~Colombi, E.~Gaztanaga and R.~Scoccimarro, \emph{{Large scale
  structure of the universe and cosmological perturbation theory}},
  \href{https://doi.org/10.1016/S0370-1573(02)00135-7}{\emph{Phys. Rept.}
  {\bfseries 367} (2002) 1}
  [\href{https://arxiv.org/abs/astro-ph/0112551}{{\ttfamily
  astro-ph/0112551}}].

\bibitem{Newburgh:2016mwi}
L.~Newburgh et~al., \emph{{HIRAX: A Probe of Dark Energy and Radio
  Transients}}, \href{https://doi.org/10.1117/12.2234286}{\emph{Proc. SPIE Int.
  Soc. Opt. Eng.} {\bfseries 9906} (2016) 99065X}
  [\href{https://arxiv.org/abs/1607.02059}{{\ttfamily 1607.02059}}].

\bibitem{Alonso:2017dgh}
D.~Alonso, P.~G. Ferreira, M.~J. Jarvis and K.~Moodley, \emph{{Calibrating
  photometric redshifts with intensity mapping observations}},
  \href{https://doi.org/10.1103/PhysRevD.96.043515}{\emph{Phys. Rev. D}
  {\bfseries 96} (2017) 043515}
  [\href{https://arxiv.org/abs/1704.01941}{{\ttfamily 1704.01941}}].

\bibitem{Zhu:2016esh}
H.-M. Zhu, U.-L. Pen, Y.~Yu and X.~Chen, \emph{{Recovering lost 21 cm radial
  modes via cosmic tidal reconstruction}},
  \href{https://doi.org/10.1103/PhysRevD.98.043511}{\emph{Phys. Rev. D}
  {\bfseries 98} (2018) 043511}
  [\href{https://arxiv.org/abs/1610.07062}{{\ttfamily 1610.07062}}].

\bibitem{Modi:2019hnu}
C.~Modi, M.~White, A.~Slosar and E.~Castorina, \emph{{Reconstructing
  large-scale structure with neutral hydrogen surveys}},
  \href{https://doi.org/10.1088/1475-7516/2019/11/023}{\emph{JCAP} {\bfseries
  11} (2019) 023} [\href{https://arxiv.org/abs/1907.02330}{{\ttfamily
  1907.02330}}].

\bibitem{Witzemann:2018cdx}
A.~Witzemann, D.~Alonso, J.~Fonseca and M.~G. Santos, \emph{{Simulated
  multitracer analyses with HI intensity mapping}},
  \href{https://doi.org/10.1093/mnras/stz778}{\emph{Mon. Not. Roy. Astron.
  Soc.} {\bfseries 485} (2019) 5519}
  [\href{https://arxiv.org/abs/1808.03093}{{\ttfamily 1808.03093}}].

\bibitem{Schmit:2018rtf}
C.~J. Schmit, A.~F. Heavens and J.~R. Pritchard, \emph{{The gravitational and
  lensing-ISW bispectrum of 21 cm radiation}},
  \href{https://doi.org/10.1093/mnras/sty3400}{\emph{Mon. Not. Roy. Astron.
  Soc.} {\bfseries 483} (2019) 4259}
  [\href{https://arxiv.org/abs/1810.00973}{{\ttfamily 1810.00973}}].

\bibitem{DiDio:2013sea}
E.~Di~Dio, F.~Montanari, R.~Durrer and J.~Lesgourgues, \emph{{Cosmological
  Parameter Estimation with Large Scale Structure Observations}},
  \href{https://doi.org/10.1088/1475-7516/2014/01/042}{\emph{JCAP} {\bfseries
  01} (2014) 042} [\href{https://arxiv.org/abs/1308.6186}{{\ttfamily
  1308.6186}}].

\bibitem{Yankelevich:2018uaz}
V.~Yankelevich and C.~Porciani, \emph{{Cosmological information in the
  redshift-space bispectrum}},
  \href{https://doi.org/10.1093/mnras/sty3143}{\emph{Mon. Not. Roy. Astron.
  Soc.} {\bfseries 483} (2019) 2078}
  [\href{https://arxiv.org/abs/1807.07076}{{\ttfamily 1807.07076}}].

\bibitem{Chang:2007xk}
T.-C. Chang, U.-L. Pen, J.~B. Peterson and P.~McDonald, \emph{{Baryon Acoustic
  Oscillation Intensity Mapping as a Test of Dark Energy}},
  \href{https://doi.org/10.1103/PhysRevLett.100.091303}{\emph{Phys. Rev. Lett.}
  {\bfseries 100} (2008) 091303}
  [\href{https://arxiv.org/abs/0709.3672}{{\ttfamily 0709.3672}}].

\bibitem{Villaescusa-Navarro:2018vsg}
F.~Villaescusa-Navarro et~al., \emph{{Ingredients for 21 cm Intensity
  Mapping}}, \href{https://doi.org/10.3847/1538-4357/aadba0}{\emph{Astrophys.
  J.} {\bfseries 866} (2018) 135}
  [\href{https://arxiv.org/abs/1804.09180}{{\ttfamily 1804.09180}}].

\bibitem{Jalilvand:2019bhk}
M.~Jalilvand, E.~Majerotto, C.~Bonvin, F.~Lacasa, M.~Kunz, W.~Naidoo et~al.,
  \emph{{New Estimator for Gravitational Lensing Using Galaxy and Intensity
  Mapping Surveys}},
  \href{https://doi.org/10.1103/PhysRevLett.124.031101}{\emph{Phys. Rev. Lett.}
  {\bfseries 124} (2020) 031101}
  [\href{https://arxiv.org/abs/1907.00071}{{\ttfamily 1907.00071}}].

\bibitem{Karagiannis:2019jjx}
D.~Karagiannis, A.~z. Slosar and M.~Liguori, \emph{{Forecasts on Primordial
  non-Gaussianity from 21 cm Intensity Mapping experiments}},
  \href{https://arxiv.org/abs/1911.03964}{{\ttfamily 1911.03964}}.

\bibitem{Ansari:2018ury}
{\scshape Cosmic Visions 21 cm} collaboration, \emph{{Inflation and Early Dark
  Energy with a Stage II Hydrogen Intensity Mapping experiment}},
  \href{https://arxiv.org/abs/1810.09572}{{\ttfamily 1810.09572}}.

\bibitem{doi:10.1021/ac60214a047}
A.~Savitzky and M.~J.~E. Golay, \emph{Smoothing and differentiation of data by
  simplified least squares procedures.},
  \href{https://doi.org/10.1021/ac60214a047}{\emph{Analytical Chemistry}
  {\bfseries 36} (1964) 1627}.

\bibitem{Montanari:2020}
F.~Montanari and S.~Camera, \emph{{Speeding up the detectability of the
  harmonic-space galaxy bispectrum}},
  \href{https://arxiv.org/abs/2008.11131}{{\ttfamily 2008.11131}}.

\bibitem{DiDio:2016gpd}
E.~Di~Dio, H.~Perrier, R.~Durrer, G.~Marozzi, A.~Moradinezhad~Dizgah,
  J.~Noreña et~al., \emph{{Non-Gaussianities due to Relativistic Corrections
  to the Observed Galaxy Bispectrum}},
  \href{https://doi.org/10.1088/1475-7516/2017/03/006}{\emph{JCAP} {\bfseries
  03} (2017) 006} [\href{https://arxiv.org/abs/1611.03720}{{\ttfamily
  1611.03720}}].

\end{thebibliography}\endgroup

\end{document}